\documentclass[12pt]{article}
\usepackage{graphicx}
\def\hybrid{\topmargin 0pt      \oddsidemargin 0pt
        \headheight 0pt \headsep 0pt
        \voffset=-0.5cm
        \textwidth 6.25in       
        \textheight 9.5in       
        \marginparwidth 0.0in
        \parskip 5pt plus 1pt   \jot = 1.5ex}
\catcode`\@=11
\def\marginnote#1{}

\newcount\hour
\newcount\minute
\newtoks\amorpm
\hour=\time\divide\hour by60
\minute=\time{\multiply\hour by60 \global\advance\minute by-\hour}
\edef\standardtime{{\ifnum\hour<12 \global\amorpm={am}%
        \else\global\amorpm={pm}\advance\hour by-12 \fi
        \ifnum\hour=0 \hour=12 \fi
        \number\hour:\ifnum\minute<10 0\fi\number\minute\the\amorpm}}
\edef\militarytime{\number\hour:\ifnum\minute<10 0\fi\number\minute}

\def\draftlabel#1{{\@bsphack\if@filesw {\let\thepage\relax
   \xdef\@gtempa{\write\@auxout{\string
      \newlabel{#1}{{\@currentlabel}{\thepage}}}}}\@gtempa
   \if@nobreak \ifvmode\nobreak\fi\fi\fi\@esphack}
        \gdef\@eqnlabel{#1}}
\def\@eqnlabel{}
\def\@vacuum{}
\def\draftmarginnote#1{\marginpar{\raggedright\scriptsize\tt#1}}
\def\draftlabel#1{{\@bsphack\if@filesw {\let\thepage\relax
   \xdef\@gtempa{\write\@auxout{\string
      \newlabel{#1}{{\@currentlabel}{\thepage}}}}}\@gtempa
   \if@nobreak \ifvmode\nobreak\fi\fi\fi\@esphack}
        \gdef\@eqnlabel{#1}}
\def\@eqnlabel{}
\def\@vacuum{}
\def\draftmarginnote#1{\marginpar{\raggedright\scriptsize\tt#1}}

\def\draft{\oddsidemargin -.5truein
        \def\@oddfoot{\sl preliminary draft \hfil
        \rm\thepage\hfil\sl\today\quad\militarytime}
        \let\@evenfoot\@oddfoot \overfullrule 3pt
        \let\label=\draftlabel
        \let\marginnote=\draftmarginnote
   \def\@eqnnum{(\theequation)\rlap{\kern\marginparsep\tt\@eqnlabel}%
\global\let\@eqnlabel\@vacuum}  }

\def\numberbysection{\@addtoreset{equation}{section}
        \def\theequation{\thesection.\arabic{equation}}}

\def\underline#1{\relax\ifmmode\@@underline#1\else
        $\@@underline{\hbox{#1}}$\relax\fi}

\def\titlepage{\@restonecolfalse\if@twocolumn\@restonecoltrue\onecolumn
     \else \newpage \fi \thispagestyle{empty}\c@page\z@
        \def\thefootnote{\fnsymbol{footnote}} }

\def\endtitlepage{\if@restonecol\twocolumn \else  \fi
        \def\thefootnote{\arabic{footnote}}
        \setcounter{footnote}{0}}  
\relax

\numberbysection
\hybrid

\newfont{\Bbb}{msbm10 scaled 1\@ptsize00}
\newfont{\Bbbb}{msbm7 scaled 1\@ptsize00}
\newcommand{\CC}{\mbox{\Bbb C}}

\newcommand{\DDD}{\raise-1pt\hbox{$\mbox{\Bbbb D}$}}

\newcommand{\UUU}{\raise-1pt\hbox{$\mbox{\Bbbb U}$}}

\newcommand{\ZZ}{\mbox{\Bbb Z}}
\newcommand{\z}{\raise-1pt\hbox{$\mbox{\Bbbb Z}$}}

\def\beq{\begin{equation}}
\def\eeq{\end{equation}}
\def\p{\partial}
\def\G{{\sf \Gamma}}

\def\normord{ {\scriptstyle {{\bullet}\atop{\bullet}}} }

\def\lbr{\Bigl <}
\def\rbr{\Bigr >}

\begin{document}
\begin{titlepage}

\title{Intertwining operators for Sklyanin
algebra and elliptic hypergeometric series}

\author{A.~Zabrodin
\thanks{Institute of Biochemical Physics,
4 Kosygina st., 119334, Moscow, Russia and ITEP, 25
B.Cheremushkinskaya, 117218, Moscow, Russia}}

\date{December 2010}
\maketitle

\begin{abstract}

Intertwining operators for
infinite-dimensional representations of the Sklyanin
algebra with spins $\ell$ and $-\ell-1$
are constructed using the technique of
intertwining vectors for elliptic $L$-operator.
They are expressed in terms of elliptic hypergeometric
series with operator argument. The intertwining
operators obtained ($W$-operators) serve as
building blocks for the elliptic $R$-matrix which intertwines
tensor product of two $L$-operators taken in
infinite-dimensional representations of the Sklyanin
algebra with arbitrary spin. The Yang-Baxter equation
for this $R$-matrix follows from
simpler equations of the star-triangle type for the
$W$-operators. A natural graphic representation
of the objects and equations involved in the construction is
used.

\end{abstract}
\vfill

\end{titlepage}

\section{Introduction}

Central to the theory of quantum integrable systems is
quantum $R$-matrix satisfying the celebrated Yang-Baxter equation.
General $R$-matrices
with additive spectral parameter are
parametrized via elliptic functions.
The simplest elliptic $R$-matrix is
\beq
\label{02}
R(\lambda )=\sum_{a=0}^{3}
\frac{\theta_{a+1}(2\lambda +\eta |\tau )}
{\theta_{a+1}(\eta |\tau )}\,
\sigma_a \otimes \sigma_a \,,
\eeq
where $\theta_a(z|\tau)$ are Jacobi $\theta$-functions,
$\sigma_a$ are Pauli matrices, and $\sigma_0$ is the unit
matrix.
This $R$-matrix is associated with the celebrated
8-vertex model solved by Baxter \cite{Baxter}, being the
matrix of local Boltzmann weights at the vertex.
The transfer matrix of this model is the generating function
of conserved quantities for
the integrable anisotropic (of $XYZ$ type)
spin-$\frac{1}{2}$ chain.
Integrable spin chains of $XYZ$-type and their
higher spin generalizations can be solved
by the generalized algebraic Bethe
ansatz \cite{FT,Takebe}.

In lattice integrable models with elliptic $R$-matrix (\ref{02}),
the algebra of local observables
is the Sklyanin algebra \cite{Skl1,Skl2}
which is a special 2-parametric deformation of the
universal enveloping algebra $U(gl(2))$. A concrete
model is defined by fixing a particular representation
of this algebra. Such representations can be realized by difference
operators.
Similar to the $sl(2)$-case
(models of $XXX$-type), the representations are labeled
by a continuous parameter which is called
{\it spin}, and for positive half-integer values of this
parameter the operators representing the Sklyanin algebra
generators are known to have a finite-dimensional invariant space.
However, we allow the
spin to take any complex value, so we are going to
work in a general infinite-dimensional representation of the
algebra of observables.

Integrable spin chains of XXX-type with infinite-dimensional representations 
of symmetry algebra at the sites were first
studied in the seminal papers \cite{Lipatov,FK} in the context of
high energy QCD, see also \cite{noncompactXXX}.
Later, lattice models with trigonometric $R$-matrix (of XXZ-type)
with non-compact quantum group symmetry were considered
\cite{noncompactXXZ}. A representation-theoretical approach
to models with elliptic $R$-matrix and
``non-compact" Sklyanin algebra symmetry
is presently not available but there is no doubt that it should
exist.

In this paper we present a direct construction of the
elliptic $R$-matrix intertwining the tensor product of two
arbitrary infinite-dimensional representations of the
Sklyanin algebra. It can be realized
as a difference operator in two variables, in general of
infinite order, so we often call this object {\it $R$-operator}
rather than $R$-matrix. Another important object is a
face type $R$-matrix related to
the $R$-operator via a functional version
of the vertex-face correspondence.
The latter $R$-matrix provides an elliptic
analog of $6j$-symbols.

Our method closely follows
the similar construction in the chiral Potts model
\cite{Korepanov,BS,KMN}
and the broken $\ZZ _N$-symmetric model \cite{HasYam,brokenZ}.
It is based on the observation that the elementary
$L$-operator is in fact a composite object built from
simpler entities called ``intertwining vectors" \cite{brokenZ,Has}.
Then the proof of the Yang-Baxter equation and other
properties of the $L$-operator can be reduced
to simple manipulations with the intertwining vectors using
basic relations between them. Remarkably, all elements of this procedure
have a nice graphic interpretation which makes them
rather clear and greatly
simplifies the arguments. It provides simultaneously a very good
illustration and an important heuristic tool.
This graphical technique
resembles both the one developed for
the Chiral Potts and
broken $\ZZ _N$-symmetric models and the one
known in the representation theory of $q$-deformed
algebras, in particular in connection with $q$-deformation
of $6j$-symbols \cite{KR88}.

However, practical realization of these ideas in the
infinite-dimensional setting is by no means obvious.
Technically, it is rather different
from what is customary in the 8-vertex model and its relatives.
Our construction goes along the lines
of our earlier work \cite{Z00} devoted to the $Q$-operator
for spin chains with infinite-dimensional representations
of the Sklyanin algebra
at each site and extensively uses such really special functions as
elliptic gamma-function and elliptic generalization
of hypergeometric series. The theory of elliptic hypergeometric
functions originated by Frenkel and Turaev
in \cite{FrTur} is now an actively
developing new branch of mathematics (see, e.g.,
\cite{Spir,Warnaar,Rosengren03}
and references therein).

The elliptic $R$-operator appears to be a composite object whose
building blocks are operators which intertwine
representations of the Sklyanin algebra with spins
$\ell$ and $-\ell -1$ (the $W$-operator). They can be
expressed through the elliptic hypergeometric series ${}_4 \omega_{3}$
with an operator argument. The kernels of the $W$-operators are
expressed through ratios of the elliptic gamma-function.
These intertwining operators were found
in our earlier paper \cite{Z00} as a by-product
of the general elliptic $Q$-operator construction.
Here we re-derive this result with the help of the
intertwining vectors using
much more direct arguments.
We also give a construction of vacuum vectors
for the elliptic $L$-operator using the graphic technique
and show how they are related to the kernel of the $W$-operator.

It should be remarked that similar results,
in one or another form,
can be found in the existing literature.
In particular, the elliptic $R$-operator has been
found \cite{DKK07} in terms of
operators which implement elementary permutations of
parameters entering the $RLL=LLR$ relation.
A solution to the star-triangle equation
built from ratios of the elliptic gamma-functions
was recently suggested in \cite{BS10}.
Some closely related matters are discussed in the
recent paper \cite{Spir10}. 
It seems to us that our approach
may be of independent interest since it emphasizes
the connection with the Sklyanin algebra
and allows one
to obtain more detailed results
in a uniform way.

The paper is organized as follows.
Section 2 contains the necessary things related to
the Sklyanin algebra, its realization by
difference operators and representations.
In section 3 we describe a space of discontinuous functions
of special form, where the Sklyanin algebra acts, and which
are identified with kernels of difference
operators.
Here we follow \cite{Z00}.
The technique of intertwining vectors developed
in Section 4 is used in
Section 5 to construct operators which intertwine
representations of the Sklyanin algebra with spins
$\ell$ and $-\ell -1$. They
appear to be the most important constituents of
the elliptic $R$-operator. In section 6 we show how
the vacuum vectors for the $L$-operator constructed
in \cite{Z00} emerge within
the approach of the present paper.
The construction of the elliptic $R$-operator
and related objects for arbitrary
spin is presented in Section 7, where the Yang-Baxter and
star-triangle relations are also discussed.
Some concluding remarks are given in Section 8.
Appendix A contains necessary information on the special
functions involved in the main part of the paper. In Appendix B
some details of the calculations with elliptic hypergeometric
series are presented.

\section{Representations of the Sklyanin algebra}

The aim of this section is to give the necessary
preliminaries on representations of the Sklyanin algebra.
We begin with a few formulas related to
the quantum $L$-operator
with elliptic dependence on the spectral parameter.

The elliptic quantum $L$-operator is
the matrix
\beq
\mbox{{\sf L}}(\lambda)=\frac{1}{2}
\left ( \begin{array}{cc}
\theta _{1}(2\lambda){\bf s}_0 +
\theta _{4}(2\lambda){\bf s}_3 &
\theta _{2}(2\lambda){\bf s}_1 +
\theta _{3}(2\lambda){\bf s}_2
\\& \\
\theta _{2}(2\lambda){\bf s}_1 -
\theta _{3}(2\lambda){\bf s}_2 &
\theta _{1}(2\lambda){\bf s}_0 -
\theta _{4}(2\lambda){\bf s}_3
\end{array} \right )
\label{L}
\eeq
with non-commutative matrix elements.
Specifically,
${\bf s}_a$ are difference operators in a
complex variable $z$:
\beq
{\bf s}_{a} = \frac{\theta _{a+1}(2z -2\ell \eta)}
{\theta _{1}(2z)}\, e^{\eta \p _z}
-\frac{\theta _{a+1}(-2z -2\ell \eta)}
{\theta _{1}(2z)}\, e^{-\eta \p _z}
\label{Sa}
\eeq
introduced by Sklyanin \cite{Skl2}.
Here $\theta_a(z)\equiv \theta_a(z|\tau)$ are
Jacobi $\theta$-functions
with the elliptic module $\tau$,
$\mbox{Im}\,\tau >0$, $\ell$ is a complex number (the spin),
and $\eta \in \CC$ is a parameter which is assumed to
belong to the fundamental parallelogram
with vertices $0$, $1$, $\tau$, $1+\tau$, and to be
incommensurate with $1, \tau$.
Definitions and transformation properties
of the $\theta$-functions are listed in Appendix A.

The four operators ${\bf s}_a$ obey the commutation
relations of the Sklyanin algebra\footnote{The standard generators
of the Sklyanin algebra \cite{Skl1}, $S_a$,
are related to ours as follows:
$S_{a}=(i)^{\delta _{a,2}}\theta _{a+1}(\eta ){\bf s}_a$.}:
\beq
\begin{array}{l}
(-1)^{\alpha +1}I_{\alpha 0}{\bf s}_{\alpha}{\bf s}_{0}=
I_{\beta \gamma}{\bf s}_{\beta}{\bf s}_{\gamma}
-I_{\gamma \beta}{\bf s}_{\gamma}{\bf s}_{\beta}\,,
\\ \\
(-1)^{\alpha +1}I_{\alpha 0}{\bf s}_0 {\bf s}_{\alpha}=
I_{\gamma \beta}{\bf s}_{\beta}{\bf s}_{\gamma}
-I_{\beta \gamma}{\bf s}_{\gamma}{\bf s}_{\beta}
\end{array}
\label{skl6}
\eeq
with the structure constants
$I_{ab}=\theta _{a+1}(0)\theta _{b+1}(2\eta)$.
Here $a,b =0, \ldots , 3$ and
$\{\alpha ,\beta , \gamma \}$ stands for any cyclic
permutation of $\{1 ,2,3\}$.
The relations of the Sklyanin algebra
are equivalent to the condition that the $\mbox{{\sf L}}$-operator
satisfies the
``$R{\sf L}{\sf L}={\sf L}{\sf L}R$"
relation with the elliptic $R$-matrix (\ref{02}).

The parameter
$\ell$ in (\ref{Sa}) is called the spin of the representation.
If necessary, we write ${\bf s}_a ={\bf s}_{a}^{(\ell )}$
or ${\sf L}^{(\ell )}(\lambda )$
to indicate the dependence on $\ell$.
When $\ell \in
\frac{1}{2} \ZZ_{+}$, these operators
have a finite-dimensional invariant subspace, namely,
the space
$\Theta_{4\ell}^{+}$ of {\it even} $\theta$-functions of
order $4\ell$ (see Appendix A). This
is the representation space of the $(2\ell +1)$-dimensional
irreducible representation (of series a))
of the Sklyanin algebra. For example,
at $\ell =\frac{1}{2}$ the functions
$\bar \theta_4 (z)$, $\bar \theta_3 (z)$
(hereafter we use the notation $\bar \theta_a (z)\equiv
\theta _a (z|\frac{\tau}{2})$)
form a basis
in $\Theta_{2}^{+}$, and the generators
${\bf s}_a$, with respect to this basis,
are represented by $2\times 2$
matrices $(-i)^{\delta_{a,2}}
(\theta_{a+1}(\eta ))^{-1}\sigma_a$. In this case,
${\sf L}(\lambda )=R(\lambda -\frac{1}{2}\eta )$,
where $R$ is the 8-vertex model $R$-matrix (\ref{02}).
In general, the representation space of the Sklyanin
algebra where the operators ${\bf s}_a$ act is called
{\it quantum space} while the two-dimensional space
in which the $L$-operator is the 2$\times$2 matrix
is called {\it auxiliary space}.

As is proved in \cite{z},
the space $\Theta_{4\ell}^{+}$ for
$\ell \in \frac{1}{2}\ZZ_{+}$ is annihilated by the operator
\beq
\label{W}
{\bf W}_{\ell} = c\sum_{k=0}^{2\ell +1} (-1)^k
\left [ \begin{array}{c} 2\ell +1 \\ k \end{array} \right ]
\, \frac{ \theta _{1}(2z+
2(2\ell -2k +1)\eta )}{\prod_{j=0}^{2\ell +1}
\theta_1(2z+2(j-k)\eta )} \,
e^{(2\ell -2k +1)\eta \p_z }.
\eeq
where $c$ is a normalization constant to be
fixed below.
Hereafter, we use the
``elliptic factorial" and ``elliptic binomial" notation:
\beq
\label{binom} [j]\equiv \theta_1(2j\eta)\,,
\;\;\;\;\;\; [n]!=\prod_{j=1}^{n}[j]\,,
\;\;\;\;\;\;
\left [ \begin{array}{c}n\\m\end{array}\right ]
\equiv \displaystyle{\frac{[n]!}{[m]![n-m]!}}\,.
\eeq
The defining property of the operator ${\bf W}_{\ell}$
established in \cite{z} is that ${\bf W}_{\ell}$
intertwines representations
of spin $\ell$ and of spin $-(\ell +1)$:
\beq
\label{S3}
{\bf W}_{\ell}\,
{\bf s}_{a}^{(\ell )}=
{\bf s}_{a}^{(-\ell -1)} {\bf W}_{\ell}\,,
\;\;\;\;\;\; a=0,\ldots , 3\,.
\eeq
The same intertwining relation can be written
for the quantum $L$-operator (\ref{L}):
\beq\label{S3a}
{\bf W}_{\ell}\,
{\sf L}^{(\ell )}\, (\lambda )=
{\sf L}^{(-\ell -1)}(\lambda ) {\bf W}_{\ell}.
\eeq

Note that the operator ${\bf W}_{\ell}$ serves
as an elliptic analog of
$(d/dz)^{2\ell +1}$ in the following sense.
In the case of the algebra $sl(2)$, the intertwining
operator between representations of spins $\ell$ and
$-\ell -1$ (realized by differential operators in $z$) is
just $(d/dz)^{2\ell +1}$. It annihilates the linear space
of polynomials of degree $\leq 2\ell$ (which results in
the rational degeneration of the elliptic space
$\Theta_{4\ell}^{+}$).

For us it is very
important to note that ${\bf W}_{\ell}$ can be extended to
arbitrary complex values of $\ell$
in which case it is represented by a half-infinite series in the
shift operator $e^{2\eta \p_z}$ \cite{Z00}.
The series is an elliptic analog of the very-well-poised
basic hypergeometric series with an operator argument.
The explicit form
is given below in this paper.
The intertwining relations
(\ref{S3}) hold true in this more general case, too.

Very little is known about
infinite-dimensional representations of the Sklyanin
algebra. The difference operators (\ref{Sa})
do provide such a representation but any
characterization of the space of functions
where they are going to act is not available at the moment, at least
for continuous functions.
On the other hand, the difference character of the operators
(\ref{Sa}) suggests to consider their action on a space
of discontinuous functions of a special form. The latter are
naturally identified with kernels of difference operators.
This formalism was used in our earlier paper \cite{Z00}.
It is reviewed in the next section.

\section{Kernels of difference operators}

Let $\delta (z)$ be the function equal to zero everywhere
but at $z=0$, where it equals $1$:
$\delta (z)=0$, $z\neq 0$, $\delta (0)=1$.
(We hope that the same notation as for
the conventional delta-function will cause no confusion because
the latter will not appear in what follows.)
Clearly, $z\delta (z)=0$ and $\delta ^2 (z)=\delta (z)$.

Consider the space ${\cal C}$ of functions of the form
\beq
\label{inf1}
f(z)=\sum_{k\in\z}f_k \delta (z-\nu +2k\eta )\,,
\;\;\;\;\;\;\; f_k \in \CC \,,
\eeq
where $\nu \in \CC$.
This space is isomorphic to the direct product of $\CC$ and the
linear space of
sequences $\{ f_k \}_{k\in {\bf \z}}$. We call functions of the
form (\ref{inf1}) {\it combs}.
Clearly, the Sklyanin algebra realized as in (\ref{Sa})
acts in this space (shifting $\nu \to \nu \pm \eta$).

A comb is said to be finite from the right
(respectively, from the left)
if there exists $M\in \ZZ$ such that $f_k =0$ as $k>M$
(respectively, $k<M$). Let ${\cal C}^{\vdash}$ (respectively,
${\cal C}^{\dashv}$) be the space of combs finite from the left
(respectively, from the right).

We define the pairing
\beq
\label{inf2}
(F(z),\, \delta (z-a))=F(a)
\eeq
for any function $F(z)$, not necessarily of the form (\ref{inf1}).
In particular,
\beq
\label{inf3}
(\delta (z-a),\, \delta (z-b))=
\delta (a-b)\,.
\eeq
Formally, this pairing can be written as an integral:
\beq\label{pairing}
(F(z),\, \delta (z-a))=
\int dz F(z)\delta (z-a)
\eeq
(perhaps a $q$-integral symbol would be more appropriate).
We stress that the integral here means nothing more
than another notation for the pairing,
especially convenient in case of many variables.
By linearity, the pairing can be extended to the whole
space of combs. We note that the pairing between the spaces
${\cal C}^{\vdash}$ and ${\cal C}^{\dashv}$ is well defined
since the sum is always finite.

Combs are to be thought of as kernels of difference operators.
By a difference operator in one variable we mean any
expression of the form
\beq
\label{D1}
{\bf D}=\sum_{k\in {\z}}c_k(z)e^{(\mu +2k\eta )\p _{z}}\,,
\;\;\;\;\;\; \mu \in \CC\,.
\eeq
The comb
\beq
\label{D2}
D(z, \zeta )=
\sum_{k\in {\z}}c_k(z)
\delta (z-\zeta +\mu +2k\eta )\,,
\eeq
regarded as a function of any one of the variables $z$, $\zeta$,
is the kernel of this difference operator in the following sense.
Using the pairing introduced above, we can write:
\beq
\label{D3}
({\bf D}f)(z)=
\int D(z,\zeta )f(\zeta )d\zeta =
\sum_{k\in {\z}}c_k(z)f(z+\mu +2k\eta )\,.
\eeq
The kernel $D(z, \zeta )$ can be viewed as an infinite matrix
with continuously numbered rows ($z$) and columns ($\zeta$).
Then the convolution with respect to the second
argument of the kernel, as in (\ref{D3}),
defines action of the operator from the left.
The convolution with respect to the first argument
defines the action from the right,
\beq
\label{D4}
(f{\bf D})(z)=
\int f(\zeta )D(\zeta ,z )d\zeta \,,
\eeq
equivalent to the action of the transposed
difference operator from the left:
\beq
\label{D5}
{\bf D}^{{\sf t}}=
\sum_{k\in {\z}}
e^{-(\mu +2k\eta )\p _{z}}
c_k(z) =
\sum_{k\in {\z}}
c_k(z-\mu -2k\eta )
e^{-(\mu +2k\eta )\p _{z}}\,.
\eeq
The transposition ${\sf t}$ is the anti-automorphism of the
algebra of difference operators such that
$\bigl ( c(z)e^{\alpha \p_{z}}\bigr )^{{\sf t}}=
e^{-\alpha \p_{z}}c(z)$.
In terms of the above pairing we can write
$(f, {\bf D}g)=({\bf D}^{{\sf t}}f, g)$.

The following simple remarks
will be useful in what follows. Let $F(z), G(z)$ be any functions,
then $F(z)D(z, \zeta )G(\zeta )$,
with $D(z, \zeta )$ as above, is the kernel of the difference
operator
$$
F{\bf D}G=\sum_{k\in {\z}}c_k(z)
F(z)G(z+\mu +2k\eta ) e^{(\mu +2k\eta )\p _{z}}
$$
which is the composition of the multiplication by $G$,
action of the operator ${\bf D}$ and subsequent multiplication by $F$.
Let $D^{(1)}(z, \zeta )$, $D^{(2)}(z, \zeta )$ be kernels of
difference operators ${\bf D}^{(1)}$, ${\bf D}^{(2)}$
respectively, then the convolution
$$
\int d\xi D^{(2)}(z, \xi )D^{(1)}(\xi , \zeta )
$$
is the kernel of the difference operator ${\bf D}^{(2)}{\bf D}^{(1)}$.
If the kernels $D^{(1)}(z, \zeta )$, $D^{(2)}(z, \zeta )$
are combs finite from the left (right)
as functions of $z$, then the convolution is always well defined
and the resulting kernel belongs to the same space of combs.

The kernels of Sklyanin's operators (\ref{Sa}) are:
\beq
\label{D7}
s_a(z,z')=
\frac{\theta _{a+1}(2z -2\ell \eta)}
{\theta _{1}(2z)}\,\delta (z-z' +\eta)
-\frac{\theta _{a+1}(-2z -2\ell \eta)}
{\theta _{1}(2z)}\, \delta (z-z' -\eta )\,.
\eeq
Note that $s_a(-z,-z')=s_a(z,z')$.
Let us find the kernel of the $L$-operator (\ref{L}).
Using identities for theta-functions, it is easy to see that
\beq\label{D7a}
L_{\zeta}^{z}(\lambda )=\theta_{1}(2\lambda +2\ell \eta )
V^{-1}(\lambda  +  \ell \eta ,z)
\left ( \begin{array}{cc}
\delta (z\! -\! \zeta \! +\! \eta ) &0 \\
0 &  \delta (z\! -\! \zeta \! -\! \eta ) \end{array} \right )
V(\lambda - \ell \eta ,z),
\eeq
where $V(\lambda , z)$ is the matrix
$$
V(\lambda , z)=\left ( \begin{array}{cc}
\bar \theta_{4}(z+\lambda )&
\bar \theta_{3}(z+\lambda ) \\
\bar \theta_{4}(z-\lambda )&
\bar \theta_{3}(z-\lambda ) \end{array} \right )
$$
and $V^{-1}(\lambda , z)$ is its inverse:
$$
V^{-1}(\lambda , z)=\frac{1}{2\theta _1(2z)}
\left ( \begin{array}{rr}
\bar \theta_{3}(z-\lambda )&
-\bar \theta_{3}(z+\lambda ) \\
-\bar \theta_{4}(z-\lambda )&
\bar \theta_{4}(z+\lambda ) \end{array} \right )
$$
(recall that $\bar \theta_{a}(z)\equiv \theta_{a}(z|\frac{\tau}{2})$).
A crucial point is that the diagonal matrix with delta-functions
factorizes into the product of column and row vectors:
$$
\left ( \begin{array}{cc}
\delta (z\! -\! \zeta \! +\! \eta ) &0 \\
0 &  \delta (z\! -\! \zeta \! -\! \eta ) \end{array} \right )
=\left ( \begin{array}{c}
\delta (z\! -\! \zeta \! +\! \eta )\\
\delta (z\! -\! \zeta \! -\! \eta )
\end{array} \right )
\Bigl (\delta (z\! -\! \zeta \! +\! \eta ),
\delta (z\! -\! \zeta \! -\! \eta ) \Bigr )
$$
and thus so does $L^{z}_{\zeta}(\lambda )$.
The vectors which  represent the factorized
kernel of the $L$-operator are
``intertwining vectors" introduced
in the next section.

\section{Intertwining vectors}

We introduce the 2-component (co)vector
\beq\label{vectors}
\bigl |\zeta \bigr >=
\left (
\begin{array}{l}\bar \theta_{4}(\zeta)\\
\bar \theta_{3}(\zeta)
\end{array}
\right ),
\quad \quad
\bigl < \zeta \bigr |= \bigl (
\bar \theta_{4}(\zeta),\,
\bar \theta_{3}(\zeta) \bigr ).
\eeq
The vector orthogonal to $\bigl < \zeta \bigr |$ is
$\bigl |\zeta \bigr >^{\bot}=
\left (
\begin{array}{r}\bar \theta_{3}(\zeta)\\
-\bar \theta_{4}(\zeta)
\end{array}
\right )$,
the covector orthogonal to $\bigl | \zeta \bigr >$ is
${}^{\bot}\! \bigl <\zeta \bigr |=\bigl ( \bar \theta_{3}(\zeta),
-\bar \theta_{4}(\zeta)\bigr )$, so
$\bigl < \zeta \bigr | \zeta \bigr >^{\bot}=
{}^{\bot}\!\bigl < \zeta \bigr | \zeta \bigr >=0$.
More generally, we have:
\beq
\label{scprod}
\bigl < \xi \bigr |\zeta \bigr >^{\bot}=
2\theta_{1}(\xi +\zeta )\theta_{1}(\xi -\zeta )=-
{}^{\bot}\! \bigl < \xi \bigr |\zeta \bigr > \,.
\eeq
Note also that
\beq
\label{orth}
\bigl | \zeta +{\scriptstyle \frac{1}{2} }
(1+\tau )\bigr >=
e^{-\frac{\pi i \tau}{2} -2\pi i \zeta}
\bigl | \zeta \bigr >^{\bot}.
\eeq

\begin{figure}[t]
   \centering
        \includegraphics[angle=-00,scale=0.50]{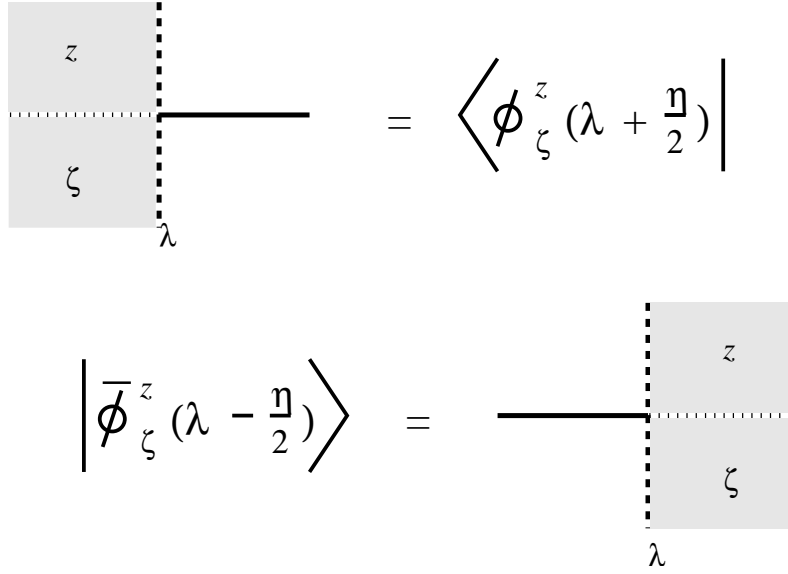}
        \caption{\it Intertwining vectors.}
    \label{fig:intwvec}
\end{figure}

Introduce now the {\it intertwining vectors}
\beq\label{intw1}
\Bigl | \phi_{z'}^{z}(\lambda )\rbr  =
\frac{1}{\sqrt{2\theta _1 (2z)}}\Bigl (
\bigl |z+\lambda \bigr > \delta (z-z'+\eta )
+\bigl |z-\lambda \bigr > \, \delta (z-z'-\eta )\Bigr ),
\eeq
\beq\label{intw1a}
\Bigl | \bar \phi_{z'}^{z}(\lambda )\rbr =
\frac{1}{\sqrt{2\theta _1 (2z)}}\Bigl (
\bigl |z-\lambda \bigr >^{\bot} \delta (z-z'+\eta )
-\bigl |z+\lambda \bigr >^{\bot}\delta (z-z'-\eta )\Bigr )
\eeq
and the corresponding covectors
\beq\label{intw2}
\Bigl < \phi_{z'}^{z}(\lambda )\Bigr |  =
\frac{1}{\sqrt{2\theta _1 (2z)}}\Bigl (
\bigl < z+\lambda \bigr | \delta (z-z'+\eta )
+\bigl <z-\lambda \bigr | \, \delta (z-z'-\eta )\Bigr ),
\eeq
\beq\label{intw2a}
\Bigl < \bar \phi_{z'}^{z}(\lambda )\Bigr | =
\frac{1}{\sqrt{2\theta _1 (2z)}}\Bigl (
{}^{\bot}\! \bigl < z-\lambda \bigr | \delta (z-z'+\eta )
-{}^{\bot} \! \bigl <z+\lambda \bigr |\delta (z-z'-\eta )\Bigr ).
\eeq
It is easy to check that
$$
\Bigl | \phi_{z'}^{z}(-\lambda )\rbr
=\sqrt{\frac{\theta_1 (2z')}{\theta_1 (2z)}}
\,\,\Bigl |\phi_{z}^{z'}(\lambda +\eta )\rbr \,,
$$
$$
\Bigl | \bar \phi_{z'}^{z}(-\lambda )\rbr
=-\sqrt{\frac{\theta_1 (2z')}{\theta_1 (2z)}}
\,\,\Bigl | \bar \phi_{z}^{z'}(\lambda -\eta )\rbr \,.
$$
The intertwining vectors satisfy the following orthogonality
relations:
\beq\label{orth1}
\lbr \phi_{z'}^{z}(\lambda )\Bigr | \,
\bar \phi_{z ''}^{z}(\lambda )\rbr =
\theta_1(2\lambda )\delta (z'-z'') \Bigl (
\delta (z\! - \! z'\! +\! \eta )+
\delta (z\! - \! z'\! -\! \eta )\Bigr ),
\eeq
\beq\label{orth2}
\lbr \phi_{z}^{z'}(\lambda +\eta )\Bigr | \,
\bar \phi_{z}^{z''}(\lambda -\eta )\rbr =
\theta_1(2\lambda )\frac{\theta_1(2z)}{\theta_1(2z')}
\, \delta (z'-z'') \Bigl (
\delta (z\! - \! z'\! +\! \eta )+
\delta (z\! - \! z'\! -\! \eta )\Bigr ),
\eeq
\beq\label{orth3}
\int d\zeta \, \Bigr | \,
\bar \phi_{\zeta}^{z}(\lambda )\rbr
\lbr \phi_{\zeta}^{z}(\lambda )\Bigr | =
\theta_1(2\lambda )
\left (\begin{array}{cc}1 & 0 \\ 0& 1 \end{array}\right ),
\eeq
\beq\label{orth4}
\int d\zeta \, \frac{\theta_1(2\zeta )}{\theta_1(2z)}\,
\Bigr | \, \bar \phi_{z}^{\zeta}(\lambda -\eta )\rbr
\lbr \phi_{z}^{\zeta}(\lambda +\eta )\Bigr | =
\theta_1(2\lambda )
\left (\begin{array}{cc}1 & 0 \\ 0& 1 \end{array}\right ).
\eeq

\begin{figure}[t]
   \centering
        \includegraphics[angle=-00,scale=0.45]{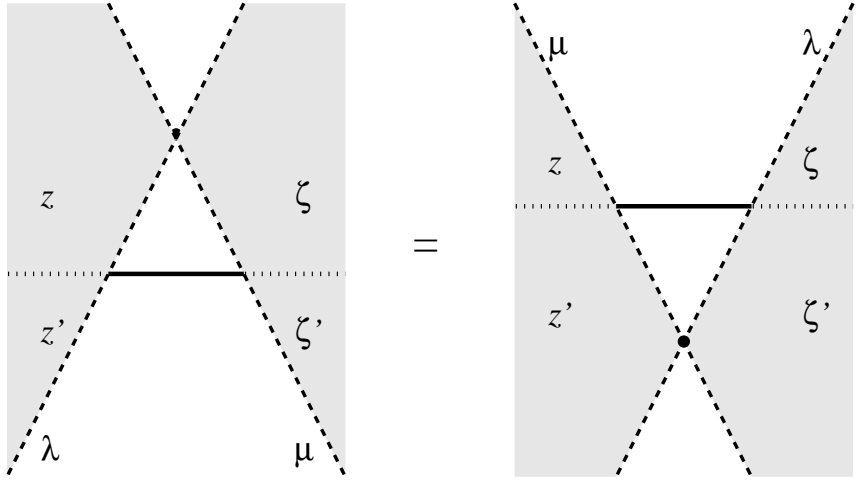}
        \caption{\it The graphic representation of the relation
        $W^{z,\zeta}(\lambda - \mu)
\lbr \phi_{z'}^{z}(\lambda +\frac{\eta}{2})\Bigr | \,
\bar \phi_{\zeta '}^{\zeta}(\mu - \frac{\eta}{2})\rbr
=W^{z',\zeta '}(\lambda -\mu )
\lbr \phi_{z'}^{z}(\mu+\frac{\eta}{2})\Bigr | \,
\bar \phi_{\zeta '}^{\zeta}(\lambda -\frac{\eta}{2} )\rbr .$
The horizontal bold line segment common for the covector
to the left and the vector to the right means taking scalar
product of the two-dimensional (co)vectors.
The intersection point of the spectral parameter lines corresponds to
the ``vertex" $W^{z,\zeta}(\lambda - \mu)$.}
    \label{fig:scalarprod}
\end{figure}

The general scalar product of two intertwining vectors is
$$
\begin{array}{lll}
\lbr \phi_{z'}^{z}(\lambda )\Bigr | \,
\bar \phi_{\zeta '}^{\zeta}(\mu )\rbr \!\! & = &
\displaystyle{
\frac{1}{\sqrt{4\theta_1(2z)\theta_1(2\zeta )}}}
\\ && \\
&\times &\Bigl \{\theta_1(z\! +\! \zeta \! +\! \lambda \! -\! \mu )
\theta_1(z\! -\! \zeta \! +\! \lambda \! +\! \mu )
\delta (z\! - \! z'\! +\! \eta )
\delta (\zeta \! - \! \zeta '\! +\! \eta )
\\ && \\
&& -\,\,
\theta_1(z\! +\! \zeta \! +\! \lambda \! +\! \mu )
\theta_1(z\! -\! \zeta \! +\! \lambda \! -\! \mu )
\delta (z\! - \! z'\! +\! \eta )
\delta (\zeta \! - \! \zeta '\! -\! \eta )
\\ && \\
&& +\,\,
\theta_1(z\! +\! \zeta \! -\! \lambda \! -\! \mu )
\theta_1(z\! -\! \zeta \! -\! \lambda \! +\! \mu )
\delta (z\! - \! z'\! -\! \eta )
\delta (\zeta \! - \! \zeta '\! +\! \eta )
\\ && \\
&& -\,\,
\theta_1(z\! +\! \zeta \! -\! \lambda \! +\! \mu )
\theta_1(z\! -\! \zeta \! -\! \lambda \! -\! \mu )
\delta (z\! - \! z'\! -\! \eta )
\delta (\zeta \! - \! \zeta '\! -\! \eta )\Bigr \} .
\end{array}
$$
It is a matter of direct verification to see that such scalar
products satisfy the ``intertwining relation":
\beq\label{intw3}
W^{z,\zeta}(\lambda - \mu)
\lbr \phi_{z'}^{z}(\lambda +\frac{\eta}{2})\Bigr | \,
\bar \phi_{\zeta '}^{\zeta}(\mu - \frac{\eta}{2})\rbr
=W^{z',\zeta '}(\lambda -\mu )
\lbr \phi_{z'}^{z}(\mu+\frac{\eta}{2})\Bigr | \,
\bar \phi_{\zeta '}^{\zeta}(\lambda -\frac{\eta}{2} )\rbr \,,
\eeq
where the quantities $W^{z,\zeta}(\lambda )$ solve
the following difference equations in $z,\zeta $:
\beq\label{intw3a}
\begin{array}{l}
\displaystyle{
W^{z+\eta, \zeta +\eta}(\lambda )
=\frac{\theta_1 (z+\zeta +\lambda +\eta )}{\theta_1
(z+\zeta +\lambda +\eta )}\, W^{z, \zeta}(\lambda )},
\\ \\
\displaystyle{
W^{z+\eta, \zeta -\eta}(\lambda )
=\frac{\theta_1 (z-\zeta +\lambda +\eta )}{\theta_1
(z+\zeta +\lambda +\eta )}\, W^{z, \zeta}(\lambda )}.
\end{array}
\eeq
These equations can be solved in terms of the elliptic
gamma-function $\Gamma (z|\tau , 2\eta):=\G (z)$
\cite{R3,FV3}
(see Appendix A):
\beq\label{intw4}
W^{z, \zeta}(\lambda )=e^{-2\pi i \lambda z/\eta}\,\,
\frac{\G (z+\zeta +\lambda +\eta )
\G (z-\zeta +\lambda +\eta )}{\G (z+\zeta -\lambda +\eta )
\G (z-\zeta -\lambda +\eta )}\,.
\eeq
There is a freedom to multiply the solution by
an arbitrary $2\eta$-periodic function of
$z+\zeta$ and $z-\zeta$. We put this function
equal to $1$. (However, this does not mean that this is the best
normalization; other possibilities will be discussed elsewhere.)
In our normalization
\beq\label{WW}
W^{z, \zeta}(\lambda )W^{z, \zeta}(-\lambda )=1
\eeq
but $W^{z, \zeta}(\lambda )$ is not symmetric under
permutation of $z$ and $\zeta$.

\begin{figure}[t]
   \centering
        \includegraphics[angle=-00,scale=0.45]{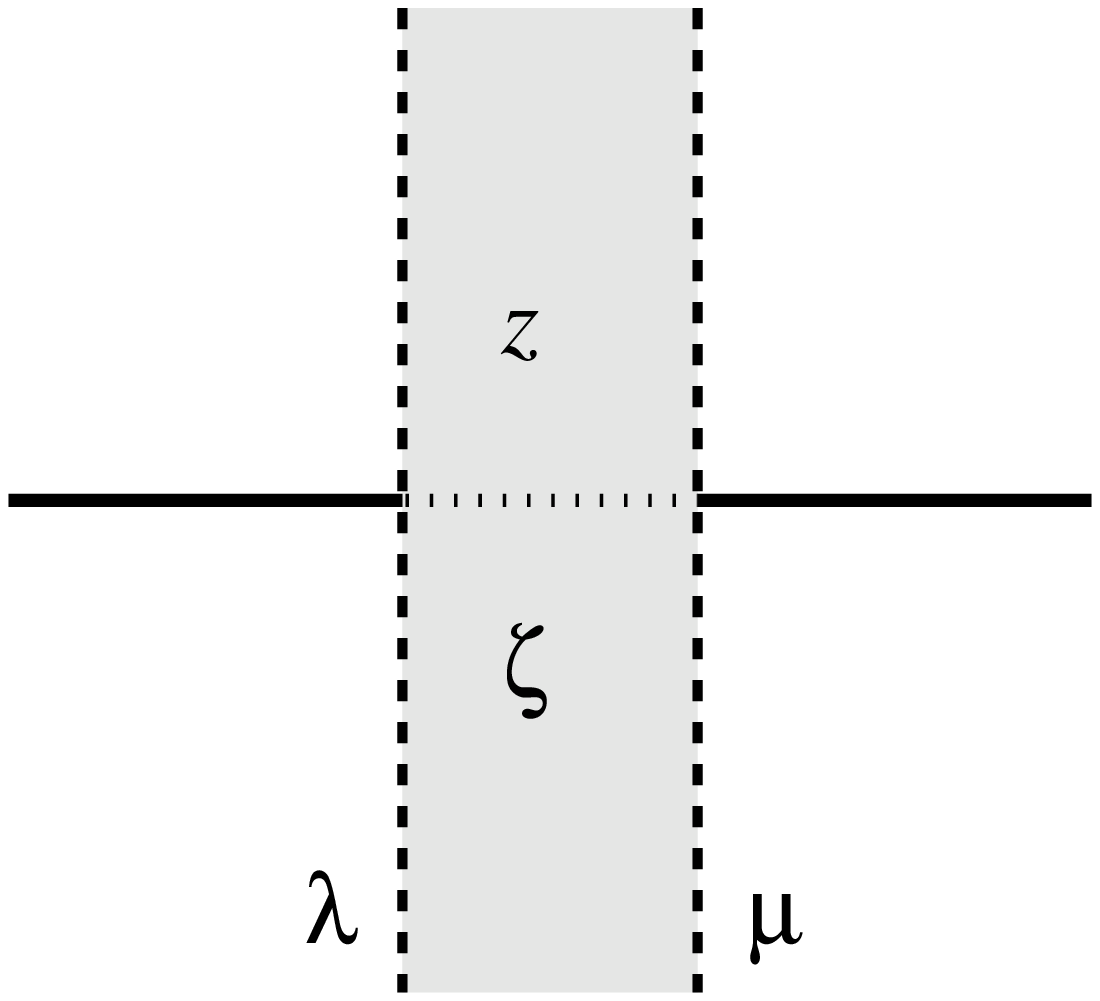}
        \caption{\it The kernel of the $L$-operator
        $L^{z}_{\zeta }(\lambda , \mu )=
        \Bigr | \, \bar \phi_{\zeta}^{z}(\lambda -
        \frac{\eta}{2} )\rbr
\lbr \phi_{\zeta}^{z}(\mu +\frac{\eta}{2})\Bigr |$.}
    \label{fig:L}
\end{figure}

The intertwining vectors can be represented graphically
as shown in Fig. \ref{fig:intwvec}. The vertical line carries the
spectral parameter and serves as a line of demarcation
between the ``real" (transparent) world and the ``shadow" world.
Then the relation (\ref{intw3})
means that the horizontal line in Fig. \ref{fig:scalarprod}
can be moved through the intersection point of the two
spectral parameter lines. This intersection point is a new
graphic element which corresponds to
$W^{z, \zeta}(\lambda -\mu )$.

The kernel of the $L$-operator for the
representation of spin $\ell$ can be written
in the factorized form as the product of intertwining
vectors:
\beq\label{intw5}
L^{\!(\ell )} {}^{z}_{\zeta }(\lambda )=
\Bigr | \, \bar \phi_{\zeta}^{z}(\lambda_{+} -\frac{\eta}{2} )\rbr
\lbr \phi_{\zeta}^{z}(\lambda_{-}+\frac{\eta}{2})\Bigr |\,,
\quad \quad \lambda_{\pm}=\lambda \pm (\ell +\frac{1}{2})\eta \,.
\eeq
It clear that the spectral parameter $\lambda$
and the representation parameter $\ell \eta$
enter here on equal footing, so the
notation $L^{\!(\ell )} {}^{z}_{\zeta }(\lambda )=
L^{z}_{\zeta }(\lambda _{+}, \lambda_{-})$
is sometimes also convenient. Graphically, the
kernel of the $L$-operator is shown in Fig.
\ref{fig:L}.

\section{Intertwining operators for arbitrary spin}

\begin{figure}[t]
   \centering
        \includegraphics[angle=-00,scale=0.45]{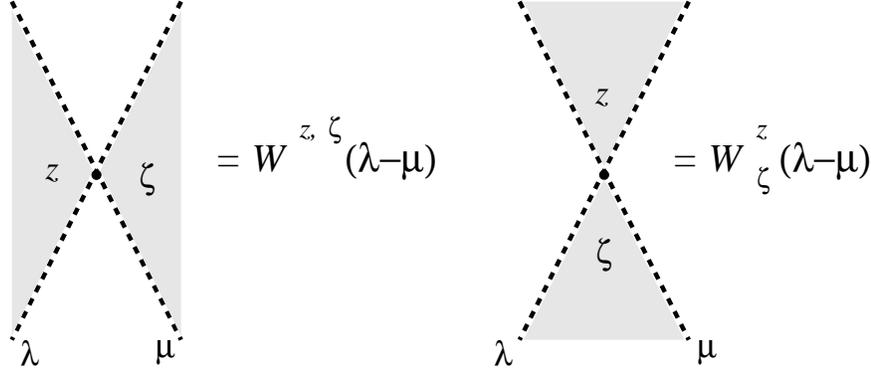}
        \caption{\it The vertices $W^{z, \zeta}(\lambda -\mu )$
        and $W^{z}_{\zeta}(\lambda -\mu )$.}
    \label{fig:vertices}
\end{figure}

\begin{figure}[t]
   \centering
        \includegraphics[angle=-00,scale=0.45]{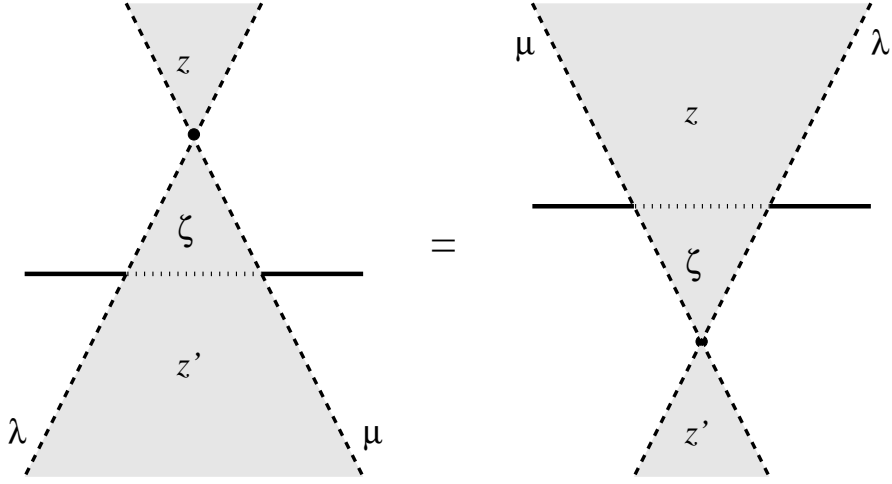}
        \caption{\it The graphical representation of equation
        (\ref{intw6}).}
    \label{fig:dualscalarprod}
\end{figure}

There is a relation which is ``dual" to (\ref{intw3})
(see also Fig. \ref{fig:scalarprod}) meaning that it can be
read from the same configuration of lines in the figure
by exchanging the real and shadow world pieces of the plane
(see Fig. \ref{fig:dualscalarprod}).
Two new elements appear: first, the
vertex $W_{\zeta}^{z}(\lambda -\mu )$ is different from the one
in Fig. \ref{fig:scalarprod} and, second, one should take
convolution ($\int d\zeta$) with respect to the ``intermediate"
variable $\zeta$ associated to the finite triangle in the
shadow world. The two vertices, $W^{z, \zeta}(\lambda -\mu )$ and
$W_{\zeta}^{z}(\lambda -\mu )$, are shown separately in
Fig. \ref{fig:vertices}.
According to Fig. \ref{fig:dualscalarprod},
the dual relation has the form
\beq\label{intw6}
\int \! d\zeta  \, W_{\zeta}^{z}(\lambda -\mu )
\Bigr | \, \bar \phi_{z'}^{\zeta}(\lambda -\frac{\eta}{2} )\rbr
\lbr \phi_{z'}^{\zeta}(\mu +\frac{\eta}{2})\Bigr | =\!
\int \! d\zeta  \, W_{z'}^{\zeta}(\lambda -\mu )
\Bigr | \, \bar \phi_{\zeta}^{z}(\mu -\frac{\eta}{2} )\rbr
\lbr \phi_{\zeta}^{z}(\lambda +\frac{\eta}{2})\Bigr |.
\eeq
Changing the notation $\lambda \to \lambda_{+}$,
$\mu \to \lambda_{-}$, one can write it as
$$
\int \! d\zeta  \, W_{\zeta}^{z}(\lambda_{+}-\lambda_{-})
L_{z'}^{\zeta}(\lambda_{+}, \lambda_{-}) =\!
\int \! d\zeta  \, W_{\zeta}^{z}(\lambda_{+}-\lambda_{-})
L_{\zeta}^{z}(\lambda_{-}, \lambda_{+})
$$
which is just the intertwining relation for the
$L$-operator ${\sf L}^{(\ell )}(\lambda )=
{\sf L}(\lambda_{+}, \lambda_{-})$ (\ref{S3a}),
with $W_{\zeta}^{z}(\lambda_{+}-\lambda_{-})$ being the
kernel of the difference operator ${\bf W}_{\ell}$.
Taking this into account, we are going to find solutions
for the $W_{\zeta}^{z}$ in the space of combs finite
either from the right or from the left.

Let us take the scalar product of both sides of equation
(\ref{intw6}) with the covector
$\lbr \phi_{z'}^{z''}(\lambda +\frac{3\eta}{2})\Bigr |$
from the left and the vector
$\Bigl |\bar \phi_{\zeta '}^{z}(\lambda +\frac{\eta}{2})\rbr$
from the right. Using the orthogonality relations
(\ref{orth1}), (\ref{orth2}), we obtain:
\beq\label{intw7}
\frac{W_{z''}^{z}(\lambda -\mu )}{\theta_1(2z'')}
\lbr \phi_{z'}^{z''}(\mu +\frac{\eta}{2})\Bigr | \,
\bar \phi_{\zeta '}^{z}(\lambda + \frac{\eta}{2})\rbr =
\frac{W_{z'}^{\zeta '}(\lambda -\mu )}{\theta_1(2z')}
\lbr \phi_{z'}^{z''}(\lambda +\frac{3\eta}{2})\Bigr | \,
\bar \phi_{\zeta '}^{z}(\mu - \frac{\eta}{2})\rbr .
\eeq
This functional relation for $W_{\zeta}^{z}$
can be solved in terms of $W^{z, \zeta}$ with the help of
(\ref{intw3}): $W_{\zeta}^{z}(\lambda )
W^{\zeta , z}(\lambda +\eta )=\theta_1 (2\zeta )$.
However, this solution is not exactly what we need
because it is not a comb-like function. Proceeding
in a slightly different way, one can rewrite (\ref{intw7})
as a system of difference equations for $W_{\zeta}^{z}(\lambda )$:
\beq\label{intw7a}
\begin{array}{l}
\displaystyle{
W_{\zeta +\eta}^{z+\eta}(\lambda )=
\frac{\theta_1(2\zeta +2\eta )}{\theta_1 (2\zeta )}\,
\frac{\theta_1(z+\zeta -\lambda )}{\theta_1
(z+\zeta +\lambda +2\eta )}\, W_{\zeta}^{z}(\lambda )},
\\ \\
\displaystyle{
W_{\zeta +\eta}^{z-\eta}(\lambda )=
\frac{\theta_1(2\zeta +2\eta )}{\theta_1 (2\zeta )}\,
\frac{\theta_1(\zeta -z -\lambda )}{\theta_1
(\zeta -z +\lambda +2\eta )}\, W_{\zeta}^{z}(\lambda )}.
\end{array}
\eeq
Comparing with (\ref{intw3a}), one immediately
finds a solution in the space of combs:
$$
W_{\zeta}^{z}(\lambda )=
\frac{c(\lambda )\theta_1 (2\zeta )}{W^{\zeta , z}(\lambda +\eta )}
\sum_{k\in \z}\delta (z-\zeta -\nu +2k\eta )
$$
with $W^{z, \zeta}$ given by (\ref{intw4}) and arbitrary $\nu$.
(The factor in front of the sum is also a solution
but in the space of meromorphic functions.)
The function $c(\lambda )$ introduced here
for the proper normalization is not determined from the
difference equations. It will be fixed below.
One may truncate the comb from the left choosing
$\nu =\lambda$; then the coefficients in front of
$\delta (z-\zeta -\lambda +2k\eta )$ with $k<0$ vanish
because the function
$W^{\zeta , z}(\lambda +\eta )$ has poles at
$\zeta =z -\lambda +2k\eta$, $k\leq -1$.
Another possibility is to truncate the comb from the right
choosing $\nu =-\lambda$; then the arguments of the delta-functions
at $k \leq 0$ exactly coincide with the half-infinite
lattice of zeros of
the function $W^{\zeta , z}(\lambda +\eta )$, and so one
can make the truncated comb by taking residues.
Below we use the first possibility and consider the solution
\beq\label{intw8}
\begin{array}{lll}
W_{\zeta}^{z}(\lambda )&=&\displaystyle{
\frac{c(\lambda )\theta_1 (2\zeta )}{W^{\zeta , z}(\lambda +\eta )}
\sum_{k\geq 0}\delta (z-\zeta -\lambda +2k\eta )}
\\ && \\
&=&\displaystyle{c(\lambda )\sum_{k\geq 0}\frac{\theta_1
(2z-2\lambda +4k\eta )}{W^{z-\lambda +2k\eta , z}(\lambda +\eta )}\,
\delta (z-\zeta -\lambda +2k\eta )}
\end{array}
\eeq
which is the kernel of the difference operator
$$
\begin{array}{lll}
{\bf W}(\lambda )&=&\displaystyle{c(\lambda )
\sum_{k\geq 0}\frac{\theta_1
(2z-2\lambda +4k\eta )}{W^{z-
\lambda +2k\eta , z}(\lambda +\eta )}\,
e^{(-\lambda +2k\eta )\p_z}}
\\ && \\
&=&\displaystyle{c(\lambda )\sum_{k\geq 0}
e^{2\pi i (\lambda +\eta )(z-\lambda +2k\eta )/\eta }
\theta_1 (2z-2\lambda +4k\eta )}
\\ && \\
&& \displaystyle{\,\,\,\,\,\,\,\, \times
\frac{\G (2z-2\lambda +2k\eta )
\G ( -2\lambda +2k\eta )}{\G ( 2z+2\eta +2k\eta )
\G ( 2\eta + 2k\eta )}\, e^{(-\lambda +2k\eta )\p_z}}.
\end{array}
$$
Rewriting the coefficients in terms of the elliptic
Pochhammer symbols with the help of (\ref{gam6}), (\ref{gam6a})
and extracting a common multiplier,
we obtain
$$
{\bf W}(\lambda )=
\tilde c(\lambda )
\frac{e^{2\pi i\lambda (z-\lambda )/\eta}
\G (-2\lambda )\G (2z-2\lambda +2\eta )}{\G (2\eta )\,
\G (2z+2\eta )}
\sum_{k\geq 0}\frac{[\frac{z-\lambda}{\eta}+
2k]}{[\frac{z-\lambda}{\eta}]\, [1]_k}\,
\frac{[\frac{z-\lambda}{\eta}]_k \,
[-\frac{\lambda}{\eta}]_k}{[\frac{z}{\eta}+1]_k }
\, e^{(-\lambda +2k\eta )\p_z},
$$
where $\tilde c(\lambda )=ie^{\frac{\pi i \tau}{6}}\eta _{D}(\tau )
c(\lambda )$.
The infinite sum can be written in terms of the
elliptic hypergeometric series ${}_{4}\omega_{3}$ (see Appendix A
for the definition) with operator argument:
\beq\label{intw9}
{\bf W}(\lambda )= \tilde c(\lambda )
\frac{e^{2\pi i\lambda (z-\lambda )/\eta}
\G (-2\lambda )\G (2z-2\lambda +2\eta )}{\G (2\eta )\,
\G (2z+2\eta )}\,
\normord {}_{4}\omega_{3}\left (\frac{z-\lambda}{\eta}; \,
-\frac{\lambda}{\eta}; \, e^{2\eta \p_z}\right )\! \normord
\, e^{-\lambda \p_z}.
\eeq
Here the double dots mean normal ordering such that the shift
operator $e^{2k\eta \p_z}$ is moved to the right.
By construction, this operator satisfies the intertwining
relation
\beq\label{intw10}
{\bf W}(\lambda -\mu ){\sf L}(\lambda , \mu )=
{\sf L}(\mu , \lambda ) {\bf W}(\lambda -\mu ).
\eeq

The intertwining property (\ref{intw10})
suggests that
${\bf W}(\lambda ){\bf W}(-\lambda )=\mbox{id}$ or, equivalently,
\beq\label{intw12}
\int \! d\zeta \, W^{z}_{\zeta}(\lambda )
W^{\zeta}_{z'}(-\lambda )=\delta (z-z')
\eeq
which is a shadow world analog of (\ref{WW}).
This is indeed true
provided that the function $c(\lambda )$ is fixed to be
\beq\label{cfixed}
c(\lambda )=
\frac{\rho_0 \, e^{\pi i \lambda ^2/\eta}}{\G (-2\lambda )},
\eeq
where the constant $\rho_0$ is
\beq\label{rho0a}
\rho_0 = \frac{\G (2\eta )}{ie^{\frac{\pi i \tau}{6}}\eta_D (\tau )}=
\frac{e^{\frac{\pi i}{12}(2\eta -3\tau )}}{i\eta_D (2\eta )}
\eeq
(clearly, there is still
a freedom to multiply $c(\lambda )$ by
a function $\varphi (\lambda )$ such that
$\varphi (\lambda )\varphi (-\lambda )=1$).
It should be noted that the very fact that the
product ${\bf W}(\lambda ){\bf W}(-\lambda )$
is proportional to the
identity operator is by no means obvious from
the infinite series representation
(\ref{intw9}). This fact was explicitly proved
in \cite{DKK07} with the
help of the Frenkel-Turaev summation formula.
For completeness, we present some details of this
calculation in Appendix B. It is this calculation that
allows one
to find $c(\lambda )$ explicitly.

We thus conclude that the
properly normalized intertwining
operator ${\bf W}(\lambda )$ reads
\beq\label{intw9a}
{\bf W}(\lambda )= e^{-\frac{\pi i\lambda ^2}{\eta} +
\frac{2\pi i \lambda z}{\eta}}\,
\frac{\G (2z\! -\! 2\lambda \! +\! 2\eta )}{\G (2z+2\eta )}\,
\normord {}_{4}\omega_{3}\left (\frac{z-\lambda}{\eta}; \,
-\frac{\lambda}{\eta}; \, e^{2\eta \p_z}\right )\! \normord
\, e^{-\lambda \p_z},
\eeq
or, in terms of the parameter $d\equiv 2\ell +1\in \CC$
related to the spin $\ell$ of the representation,
\beq\label{intw11}
{\bf W}_{\ell}\equiv {\bf W}(d\eta )=
e^{-\pi i d^2\eta +2\pi i dz}
\frac{\G (2z -2(d-1)\eta )}{\G (2z+2\eta )}\,
\normord {}_{4}\omega_{3}\left (\frac{z}{\eta} -d; \,
- d; \, e^{2\eta \p_z}\right )\! \normord
\, e^{-d\eta \p_z}.
\eeq

It is not difficult to see that the change of sign
$z\to -z$ transforms ${\bf W}(\lambda )$ to
another intertwining operator for the
Sklyanin algebra, which is an infinite series in shifts
in the opposite direction. (It is this latter
operator which was constructed in the paper \cite{Z00}.)
It can be obtained
within the same approach if one uses the other possibility
to truncate the comb which has been discussed above.
If $\ell \in \frac{1}{2}\ZZ _+$ (i.e., $d\in \ZZ _+$), then the
elliptic hypergeometric series is terminating and both operators
are represented by finite sums (containing $d+1$ terms).
Moreover, they coincide with each other and are explicitly given
by the formula
\beq\label{intw13}
{\bf W}_{\ell}=\left (ie^{\pi i (-\eta + \frac{\tau}{6})}
\eta _{D}(\tau )\right )^d
\sum_{k=0}^{d} (-1)^k
\left [ \begin{array}{c} d \\ k \end{array} \right ]
\, \frac{ \theta _{1}(2z-
2(d-2k)\eta )}{\prod_{j=0}^{d}
\theta_1(2z+2(k-j)\eta )} \,
e^{(-d +2k)\eta \p_z }
\eeq
which coincides with
equation (\ref{W}).

Let us conclude this section by summarizing the
graphic elements of the diagrams and rules of their
composing. The plane is divided
into ``transparent" and ``shadow" pieces
by a number of straight dashed lines in such a way that
each segment of any line is a border between pieces
of the different kind.
Each dashed line carries a spectral parameter denoted by
$\lambda$, $\mu$, etc.
There may be also bold straight lines which become
dotted when they go through shadow pieces of the plane.
Each shadow piece (bounded by dashed or dotted lines
or by infinity) carries a complex variable denoted by
$z$, $\zeta$, etc. Those which sit on infinite pieces
are fixed while those which sit on finite pieces bounded
by lines of any type should be ``integrated" in the sense
of the pairing (\ref{pairing}). The intersection points
of the dashed lines are of two types depending on
the way how the transparent and shadow parts are
adjacent to it. Correspondingly, there are
two types of vertex functions shown in Fig. \ref{fig:vertices}.
The intersection of a dashed line with a bold one
corresponds to an intertwining (co)vector as shown in
Fig. \ref{fig:intwvec}. Finite bold segments mean
taking scalar products of (co)vectors associated with their
endpoints.

\section{Vacuum vectors}

In order to make a closer contact with our earlier work
\cite{Z00}, it is useful to demonstrate how the vacuum vectors for
the $L$-operator can be constructed within the approach
developed in the previous sections. Let us recall the general
definition of the vacuum vectors.
Consider an arbitrary $L$-operator ${\sf L}$
with two-dimensional auxiliary space $\CC^2$, i.e., an arbitrary
$2\times 2$
operator-valued matrix
$$
{\sf L}=\left ( \begin{array}{cc} {\bf L}_{11}& {\bf L}_{12}\\
{\bf L}_{21}& {\bf L}_{22} \end{array} \right ).
$$
The operators ${\bf L}_{ij}$ act in
a linear space ${\cal H}$ which is called the quantum space of
the $L$-operator.
For the moment, let $\phi$, $\psi$, etc denote vectors from
$\CC^2$ and $X, X_1$, etc vectors from ${\cal H}$,
then acting by the quantum
$L$-operator on the tensor product
$X\otimes \phi$, we, generally speaking, obtain a mixed state
in the quantum space:
${\sf L}X\otimes \phi =X_1\otimes \phi_1+X_2\otimes \phi_2$.
The special case of a pure state,
\beq\label{V1}
{\sf L}X\otimes \phi =X'\otimes \psi \,,
\eeq
is of prime importance.
The relation (\ref{V1}) (in the particular case ${\cal H}\cong
\CC^2$) was the key point for Baxter in his solution of
the 8-vertex model \cite{Baxter}.
(This is what he called the ``pair-propagation through a vertex"
property.) Taking the scalar product
with the vector $\psi^{\bot}$ orthogonal to $\psi$, we get:
\beq
\label{V2}
\bigl (\psi^{\bot}{\sf L}\phi \bigr )X=0\,,
\eeq
i.e., the operator
${\bf K}=\bigl (\psi^{\bot}{\sf L}\phi \bigr )$ (acting in the
quantum space only) has a zero mode
$X\in {\cal H}$.
Suppose (\ref{V1}) (or (\ref{V2})) holds with some
vectors $\phi$, $\psi$;
then the vector $X$ is called
a {\it vacuum vector} of the $L$-operator.
An algebro-geometric
approach to the equation (\ref{V1}) for finite-dimensional
matrices ${\bf L}_{ik}$ was suggested by Krichever \cite{krivac}
and further developed in \cite{kz95,kz99}. In our paper \cite{Z00}
the Baxter's method of vacuum vectors was adopted to the
infinite-dimensional representations of the Sklyanin algebra.

For $L$-operators with elliptic spectral parameter it is
convenient to pass to the elliptic parametrization of the
components of the vectors $\phi$, $\psi$ as is given by
(\ref{vectors}). Writing ${\sf L}(\lambda )\bigl |\zeta \bigr >$
(respectively, $\bigl <\zeta \bigr |{\sf L}(\lambda)$)
we mean that the 2$\times$2 matrix ${\sf L}$ acts on the
2-component vector from the left (respectively, on the
2-component covector from the right).
Similarly, we introduce right and left vacuum vectors
$X_R$, $X_L$ according to the relations
\beq
\label{V3}
\bigl < \zeta \bigr | {\sf L}(\lambda)X_R =
\bigl < \xi \bigr | X'_R \,,
\;\;\;\;\;\;
X_L\bigl < \zeta \bigr | {\sf L}(\lambda) =
X'_L \bigl < \xi \bigr |\,.
\eeq
In the latter formula the matrix elements of
${\sf L}$ act on $X_L$ from the right.
Introducing the operator
\beq
\label{V5}
{\bf K}= {\bf K}(\zeta , \xi )=
\bigl < \zeta \bigr | {\sf L}(\lambda)\bigl |\xi \bigr >^{\bot},
\eeq
we can rewrite (\ref{V3}) as
${\bf K}X_R =X_L{\bf K}=0$.
The explicit form of the operator ${\bf K}$ can be
found from (\ref{L}),(\ref{Sa}):
\beq
\label{E1}
{\bf K}={\bf K}(\zeta, \xi)=\rho (z) e^{\eta \p_{z}}+
\rho (-z) e^{-\eta \p_{z}}\,,
\eeq
where
$$
\rho (z)=\frac{1}{\theta_{1}(2z)}
\prod_{\epsilon =\pm}\theta_{1}\Bigl (z+\epsilon \zeta
-\lambda_{+}+\frac{\eta}{2} \Bigr )
\theta_{1}\Bigl (z+\epsilon \xi
+\lambda_{-} +\frac{\eta}{2} \Bigr )\,.
$$
These difference operators appeared in \cite{kz99,Z00} and later were
independently introduced in \cite{Rosengren04,Rains}.
So, the equations for the right and left vacuum vectors
read
\beq
\label{R1a}
\rho (z)X_{R}(z+\eta)=
-\rho (-z)X_{R}(z-\eta)\,,
\eeq
\beq
\label{L1a}
\rho (-z-\eta)X_{L}(z+\eta)=
-\rho (z-\eta)X_{L}(z-\eta)\,.
\eeq

\begin{figure}[t]
   \centering
        \includegraphics[angle=-00,scale=0.50]{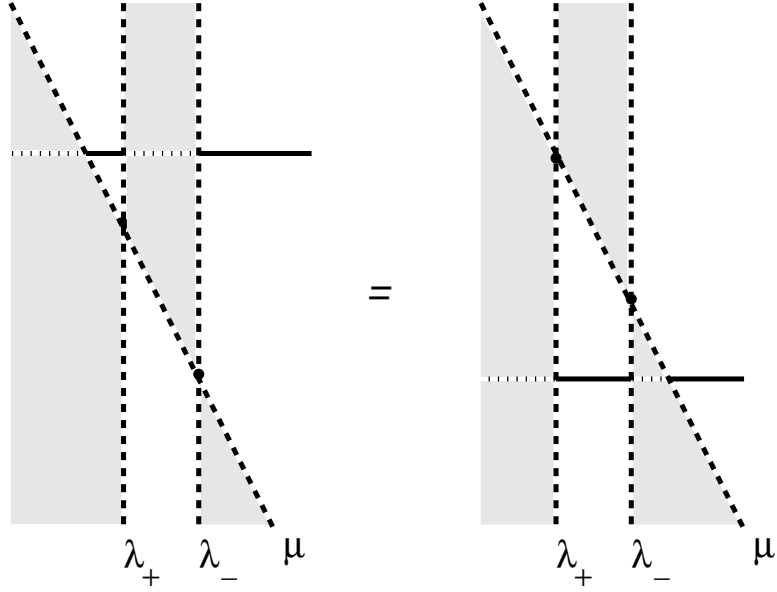}
        \caption{\it The graphic representation of equation
        (\ref{vac1}): action of the $L$-operator to right
        vacuum vectors.}
    \label{fig:vacuum}
\end{figure}

Instead of solving these equations explicitly, below we
show how the vacuum vectors emerge within the approach of the
present paper.
The key relation is
(see Fig. \ref{fig:vacuum})
\beq\label{vac1}
\begin{array}{ll}
&\displaystyle{
\int \! d\zeta \, \lbr \phi^{z'}_{\xi}(\mu +\frac{\eta}{2})
\Bigr | \bar \phi ^{z}_{\zeta}(\lambda_{+}-\frac{\eta}{2})\rbr
\lbr \phi^{z}_{\zeta}(\lambda_{-} +\frac{\eta}{2})\Bigr | \,
W^{\xi , \zeta}(\lambda_{+}-\mu)W^{\zeta}_{\xi '}(\lambda_{-}-\mu )}
\\ &\\
=&\displaystyle{
\int \! d\zeta \, \lbr \phi^{z'}_{\xi}(\lambda_{+} +\frac{\eta}{2})
\Bigr |\bar \phi ^{\zeta}_{\xi '}(\lambda_{-}-\frac{\eta}{2})\rbr
\lbr \phi^{\zeta}_{\xi '}(\mu +\frac{\eta}{2})\Bigr | \,
W^{z' , z}(\lambda_{+}-\mu)W^{z}_{\zeta}(\lambda_{-}-\mu )}.
\end{array}
\eeq
The left-hand side represents the action of the
$L$-operator ${\sf L}^{(\ell )}(\lambda )=
{\sf L}(\lambda_{+}, \lambda_{-})$ to the covector
$\lbr \phi^{z'}_{\xi}(\lambda_{+} \! +\! \frac{\eta}{2})
\Bigr |$ in the auxiliary space from the right and to the vector
$W^{\xi , \zeta}(\lambda_{+}-\mu)W^{\zeta}_{\xi '}(\lambda_{-}-\mu )$
in the quantum space from the left. It is convenient to denote
\beq\label{X}
X_{R}^{\xi, \xi '}(z|\lambda _{+}, \lambda _{-})=
W^{\xi , z}(\lambda_{+}-\frac{\eta}{2})\,
W^{z}_{\xi '}(\lambda_{-}-\frac{\eta}{2}),
\eeq
then relation (\ref{vac1}) can be rewritten (after setting
$z'=\xi \pm \eta$ and some transformations) as the system of equations
\beq\label{vac2}
\begin{array}{l}
\displaystyle{
\lbr \xi -\mu \Bigr |{\sf L}(\lambda_{+} +\mu , \lambda_{-}+\mu )
X_{R}^{\xi, \xi '}=a\lbr \xi ' -\mu \Bigr |\,
X_{R}^{\xi +\eta, \xi '+\eta}+b
\lbr \xi ' +\mu \Bigr | \,X_{R}^{\xi +\eta, \xi '-\eta}},
\\ \\
\displaystyle{
\lbr \xi +\mu \Bigr |{\sf L}(\lambda_{+}+\mu , \lambda_{-}+\mu )
X_{R}^{\xi, \xi '} =
c\lbr \xi ' -\mu \Bigr | \,X_{R}^{\xi -\eta, \xi '+\eta}
+d \lbr \xi ' +\mu \Bigr | \,X_{R}^{\xi -\eta, \xi '-\eta}},
\end{array}
\eeq
where $X_{R}^{\xi, \xi '} =X_{R}^{\xi, \xi '}
(z|\lambda_{+}, \,
\lambda_{-})$ and
$$
a=-\frac{\theta_1 (\xi \! +\! \xi '\! -\!
\lambda_{+}\! +\! \lambda_{-}\! +\! \eta )
\theta_1 (\xi \! -\! \xi '\! -\!
\lambda_{+}\! -\! \lambda_{-}\! -\! 2\mu )}{\theta_1
(2\xi ' +2\eta )}\,,
$$
$$
b=\frac{\theta_1 (\xi \! -\! \xi '\! -\!
\lambda_{+}\! +\! \lambda_{-}\! +\! \eta )
\theta_1 (\xi \! +\! \xi '\! -\!
\lambda_{+}\! -\! \lambda_{-}\! -\! 2\mu)}{\theta_1
(2\xi ' -2\eta )}\,,
$$
$$
c=-\frac{\theta_1 (\xi \! -\! \xi '\! +\!
\lambda_{+}\! -\! \lambda_{-}\! -\! \eta )
\theta_1 (\xi \! +\! \xi '\! +\!
\lambda_{+}\! +\! \lambda_{-}\! +\! 2\mu)}{\theta_1
(2\xi ' +2\eta )}\,,
$$
$$
d=\frac{\theta_1 (\xi \! +\! \xi '\! +\!
\lambda_{+}\! -\! \lambda_{-}\! -\! \eta )
\theta_1 (\xi \! -\! \xi '\! +\!
\lambda_{+}\! +\! \lambda_{-}\! +\! 2\mu)}{\theta_1 (2\xi '-2\eta )}.
$$

We note that setting $\mu =0$ one obtains from
(\ref{vac2})
$$
\lbr \xi \Bigr |{\sf L}(\lambda_{+}, \lambda_{-})
X_{R}^{\xi, \xi '}=\lbr \xi ' \Bigr |
\left (a_0 X_{R}^{\xi +\eta , \xi '+\eta}+b_0
X_{R}^{\xi +\eta , \xi '-\eta}\right )=\lbr \xi ' \Bigr |
\left (c_0 X_{R}^{\xi -\eta , \xi '+\eta}+d_0
X_{R}^{\xi -\eta , \xi '-\eta}\right ),
$$
where $a_0 =a(\mu =0)$, etc.
This means that $X_{R}^{\xi, \xi '}(z|\lambda_{+}\! -\!
\frac{\eta}{2},
\lambda_{-}\! -\! \frac{\eta}{2})$ is the right vacuum vector
for the $L$-operator (see the first equation
in (\ref{V3}). Moreover, we conclude that
\beq\label{vac3}
a_0 X_{R}^{\xi +\eta , \xi '+\eta}+b_0
X_{R}^{\xi +\eta , \xi '-\eta}=
c_0 X_{R}^{\xi -\eta , \xi '+\eta}+d_0
X_{R}^{\xi -\eta , \xi '-\eta}.
\eeq
One can that the vacuum vector is in fact a composite
object. It is a product
of two $W$-functions.
Equations (\ref{intw4}), (\ref{intw8}) together with
the 3-term identity for the Jacobi theta-function imply the relation
\beq\label{vac4}
\lbr \xi \Bigr |{\sf L}(\lambda_{+}, \lambda_{-})
X_{R}^{\xi, \xi '}\bigl (z\bigr |\lambda_{+},
\lambda_{-}\bigr )=
\theta_{1}(2\lambda_{-}+\eta )
\lbr \xi ' \Bigr |
X_{R}^{\xi, \xi '}\bigl (z\bigr |\lambda_{+}\! +\! \eta ,
\lambda_{-}\! +\! \eta \bigr )
\eeq
which is equation (4.22) from our paper \cite{Z00}
written in the slightly different notation.
The left vacuum vectors can be considered in a similar way.

\section{The $R$-operator and related objects}

\begin{figure}[t]
   \centering
        \includegraphics[angle=-00,scale=0.50]{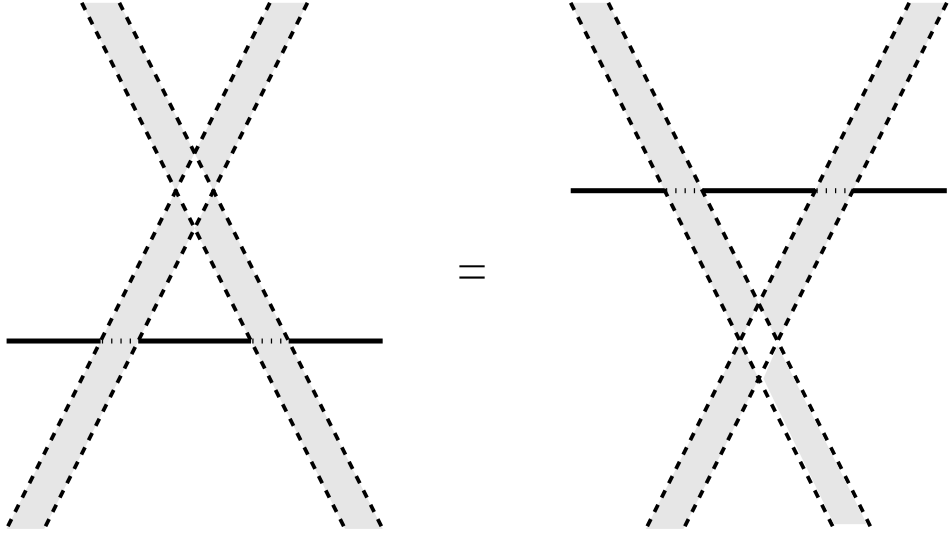}
        \caption{\it The intertwining relation
        $\check {\sf R}\, {\sf L}\otimes {\sf L}=
        {\sf L}\otimes {\sf L}\, \check {\sf R}$.}
    \label{fig:RLL}
\end{figure}

\begin{figure}[t]
   \centering
        \includegraphics[angle=-00,scale=0.50]{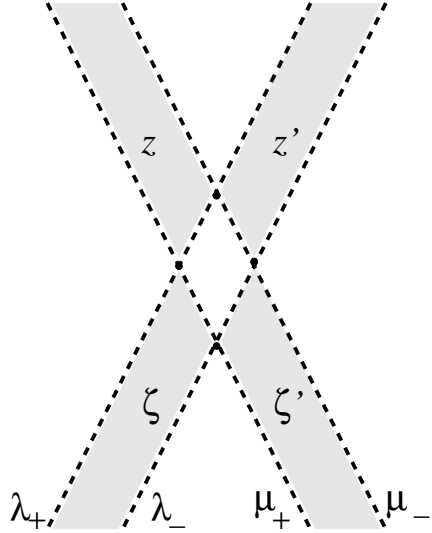}
        \caption{\it The kernel of the $R$-operator
        $R^{zz'}_{\zeta \zeta '}
(\lambda _{+}, \lambda _{-}|\mu _{+}, \mu _{-})$.}
    \label{fig:R}
\end{figure}

The $R$-operator $\check {\sf R}=
\check {\sf R}(\lambda _{+}, \lambda _{-}|
\mu _{+}, \mu _{-})$
intertwines the product of two
$L$-operators:
\beq\label{R1}
\check {\sf R}(\lambda _{+}, \lambda _{-}|
\mu _{+}, \mu _{-}){\sf L}(\lambda _{+}, \lambda _{-})
\otimes {\sf L}(\mu _{+}, \mu _{-})=
{\sf L}(\mu _{+}, \mu _{-})\otimes
{\sf L}(\lambda _{+}, \lambda _{-})
\check {\sf R}(\lambda _{+}, \lambda _{-}|
\mu _{+}, \mu _{-}).
\eeq
Passing to the different notation,
$\lambda _{\pm}=\lambda \pm (\ell +\frac{1}{2})\eta$,
$\mu _{\pm}=\mu \pm (\ell ' +\frac{1}{2})\eta$,
we can rewrite (\ref{R1}) in a more conventional form:
\beq\label{R2}
\check {\sf R}^{(\ell \ell ')}(\lambda , \mu )\,
{\sf L}^{(\ell )}(\lambda )\otimes {\sf L}^{(\ell ')}(\mu )=
{\sf L}^{(\ell ')}(\mu )\otimes
{\sf L}^{(\ell )}(\lambda )\,
\check {\sf R}^{(\ell \ell ')}(\lambda , \mu ).
\eeq
Here $\check {\sf R}^{(\ell \ell ')}(\lambda , \mu )=
\check {\sf R}(\lambda _{+}, \lambda _{-}|
\mu _{+}, \mu _{-})$ is a difference operator in
two variables acting in the tensor product of
the quantum spaces for the two $L$-operators.
In terms of the kernels equation (\ref{R1}) reads:
\beq\label{R3}
\begin{array}{ll}
&\displaystyle{\int \!  d\zeta \int \! d\zeta '\,
R^{zz'}_{\zeta \zeta '}
(\lambda _{+}, \lambda _{-}|\mu _{+}, \mu _{-})
L^{\zeta}_{\xi}(\lambda _{+}, \lambda _{-})
L^{\zeta '}_{\xi '}(\mu _{+}, \mu _{-})}
\\ & \\
=&\displaystyle{\int \!  d\zeta \int \! d\zeta '\,
L^{z}_{\zeta}(\mu _{+}, \mu _{-})
L^{z'}_{\zeta '}(\lambda _{+}, \lambda _{-})
R^{\zeta \zeta '}_{\xi \xi '}
(\lambda _{+}, \lambda _{-}|\mu _{+}, \mu _{-})}.
\end{array}
\eeq
Graphically it is shown in Fig. \ref{fig:RLL}.
The figure clarifies the structure of the kernel
of the $R$-operator which is shown in more detail
in Fig. \ref{fig:R}. It is clear that the kernel is
the product of four $W$-vertices: two of them are of
the $W^{z, \zeta}$-type (meromorphic functions) and
the other two are of the $W_{\zeta}^{z}$-type
(comb-like functions). Specifically, we can write:
$$
\begin{array}{ll}
&R^{zz'}_{\zeta \zeta '}
(\lambda _{+}, \lambda _{-}|\mu _{+}, \mu _{-})
\\ &\\
=&W^{z,z'}(\lambda _{+}-\mu_{-})
W_{\zeta '}^{z'}(\lambda_{-}-\mu_{-})
W_{\zeta}^{z}(\lambda_{+}-\mu_{+})
W^{\zeta ,\zeta '}(\lambda _{-}-\mu_{+})
\\ &\\
=&\displaystyle{W^{z,z'}(\lambda _{+}-\mu_{-})
\left [
\frac{c(\lambda _{-}-\mu _{-})
\theta_1(2\zeta ')}{W^{\zeta ',z'}(\lambda _{-}-
\mu_{-}+\eta )}\sum_{k'\geq 0}\delta (z'-\zeta ' -\lambda_{-}
+\mu_{-}+2k'\eta )\right ]}
\\ &\\
& \,\,\,\, \times \displaystyle{\left [
\frac{c(\lambda _{+}-\mu _{+})\theta_1(2\zeta )}{W^{\zeta ,z}
(\lambda _{+}-\mu_{+}+\eta )}
\sum_{k\geq 0}\delta (z-\zeta -\lambda_{+}
+\mu_{+}+2k\eta )\right ]
W^{\zeta ,\zeta '}(\lambda _{-}-\mu_{+})}
\end{array}
$$
which is the kernel of the difference
operator
\beq\label{R4a}
\check {\sf R}(\lambda _{+}, \lambda _{-}|\mu _{+}, \mu _{-})=
W^{z,z'}(\lambda_{+}-\mu_{-})
{\bf W}^{(z')}(\lambda_{-}\! -\! \mu_{-})
{\bf W}^{(z)}(\lambda_{+}\! -\! \mu_{+})
W^{z,z'}(\lambda_{-}-\mu_{+})
\eeq
(here the notation ${\bf W}^{(z)}$ means that the
operator ${\bf W}$ acts to the variable $z$.
In full, the $R$-operator reads
\beq\label{R4}
\begin{array}{ll}
&\check {\sf R}(\lambda _{+}, \lambda _{-}|\mu _{+}, \mu _{-})\, =\,
e^{-\frac{\pi i}{\eta}(\lambda_{+}-\mu _{+})^2
-\frac{\pi i}{\eta}(\lambda_{-}-\mu _{-})^2
+\frac{2\pi i}{\eta}(\lambda_{+}-\mu _{+})z
+\frac{2\pi i}{\eta}(\lambda_{-}-\mu_{-})z'}
\\ &\\
\times &\, \displaystyle{
e^{-\frac{2\pi i}{\eta}(\lambda_{+}-\mu_{-})z}
\, \frac{\G (z+z' +\lambda_{+} -\mu_{-}+\eta )
\G (z-z' +\lambda_{+}
-\mu_{-}+\eta )}{\G (z+z' -\lambda_{+} +\mu_{-}+\eta )
\G (z-z' -\lambda_{+} +\mu_{-}+\eta )}}
\\ &\\
\times &\,  \displaystyle{
\frac{\G (2z' \! -\! 2(\lambda _{-}\!
-\! \mu _{-}) \! +\! 2\eta )}{\G (2z' +2\eta )}
\, \normord {}_{4}\omega_{3}\left (
\frac{z'+\mu _{-} -\lambda _{-}}{\eta};\,
\frac{\mu _{-}-\lambda _{-}}{\eta};\,
e^{2\eta \p_{z'}}\right )\! \normord \,
e^{-(\lambda_{-}-\mu_{-})\p_{z'}}}
\\ &\\
\times &\,   \displaystyle{
\frac{\G (2z\! -\! 2(\lambda_{+}\! -\!
\mu_{+})\! +\! 2\eta )}{\G (2z +2\eta )}
\, \normord {}_{4}\omega_{3}\left (
\frac{z+\mu _{+}-\lambda _{+}}{\eta};\,
\frac{\mu _{+}-\lambda _{+}}{\eta};\,
e^{2\eta \p_{z}}\right )\! \normord \,
e^{-(\lambda_{+}-\mu_{+})\p_{z}}}
\\ &\\
\times &\,
\displaystyle{e^{-\frac{2\pi i}{\eta}(\lambda_{-}-\mu_{+})z}
\, \frac{\G (z+z' +\lambda_{-} -\mu_{+}+\eta )
\G (z-z' +\lambda_{-}
-\mu_{+}+\eta )}{\G (z+z' -\lambda_{-} +\mu_{+}+\eta )
\G (z-z' -\lambda_{-} +\mu_{+}+\eta )}}.
\end{array}
\eeq
The difference operators in the third and the fourth lines
of the r.h.s. commute because they act in different variables
but both of them do not commute with the operator of
multiplication by the function
$W^{z ,z'}(\lambda _{-}-\mu_{+})$. Note that the $R$-operator
can be also written in terms of the ${}_{6}\omega_{5}$ series
due to the identity
\beq\label{R4id}
\begin{array}{ll}
&\displaystyle{\normord
{}_{4}\omega_{3}\left (
\frac{z-\lambda}{\eta}; \, -\frac{\lambda}{\eta};\,
e^{2\eta \p_z}\right ) \normord e^{-\lambda \p_z}
W^{\zeta , z}(\mu )}
\\ & \\
=&\displaystyle{W^{\zeta , z-\lambda}(\mu )
\normord
{}_{6}\omega_{5}\left (
\frac{z-\lambda}{\eta}; \, -\frac{\lambda}{\eta},\,
\frac{z+\zeta +\mu -\lambda +\eta}{2\eta}, \,
\frac{z-\zeta +\mu -\lambda +\eta}{2\eta};\,
e^{2\eta \p_z}\right ) \normord e^{-\lambda \p_z}}.
\end{array}
\eeq

\begin{figure}[t]
   \centering
        \includegraphics[angle=-00,scale=0.50]{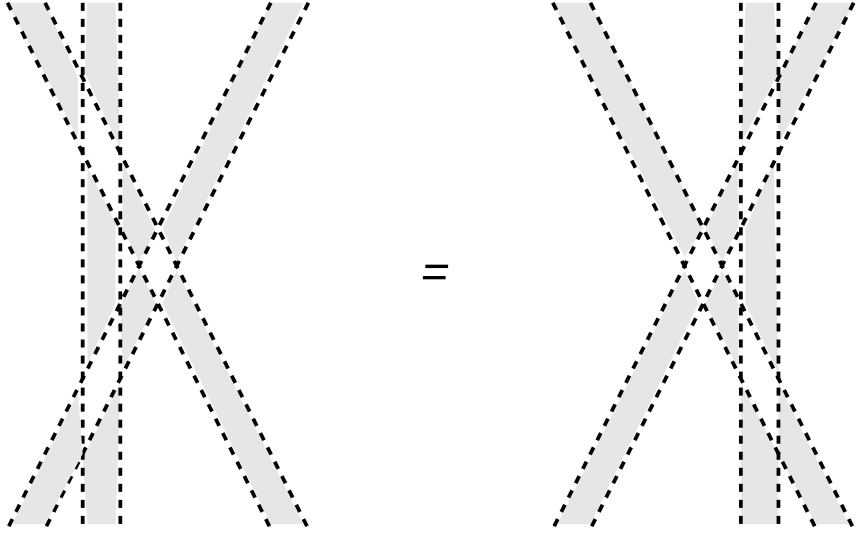}
        \caption{\it The Yang-Baxter equation for the
        $R$-operator.}
    \label{fig:RRR}
\end{figure}

\begin{figure}[t]
   \centering
        \includegraphics[angle=-00,scale=0.50]{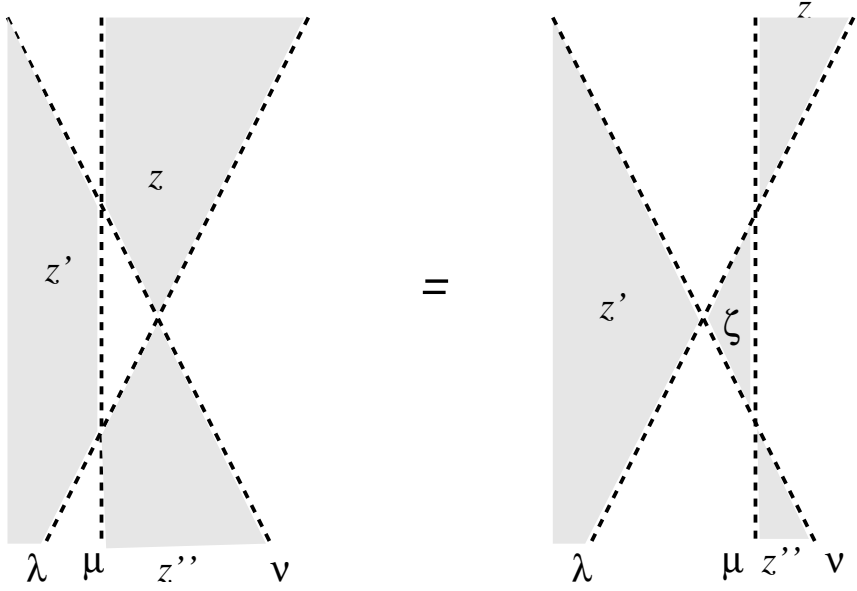}
        \caption{\it The star-triangle equation (\ref{st1a}) for the
        $W$-operators. Equation (\ref{st1b}) corresponds to the same
        configuration of lines with complimentary shadow parts
        of the plane.}
    \label{fig:WWW}
\end{figure}

The Yang-Baxter equation for the $R$-operator is schematically
shown in the self-\-exp\-la\-na\-to\-ry Fig. \ref{fig:RRR}.
One can see that
as soon as the $R$-operator is a composite object,
the Yang-Baxter equation can be reduced to simpler equations
for its elementary constituents. The latter are the
$W$-vertices of the two types. For them one can prove
a sort of the star-triangle relations
\beq\label{st1a}
W^{z',z}(\mu -\nu )W^{z', z''}(\lambda -\mu )
W^{z}_{z''}(\lambda -\nu )=
\int \! d\zeta W^{z}_{\zeta}(\lambda -\mu )
W^{z',\zeta}(\lambda -\nu )W^{\zeta}_{z''}(\mu -\nu )
\eeq
\beq\label{st1b}
W^{z,z'}(\lambda -\mu )W^{z'', z'}(\mu -\nu )
W^{z}_{z''}(\lambda -\nu )=
\int \! d\zeta W^{z}_{\zeta}(\mu -\nu )
W^{\zeta , z'}(\lambda -\nu )W^{\zeta}_{z''}(\lambda -\mu )
\eeq
schematically shown
in Fig. \ref{fig:WWW}. The proof is given in Appendix B.
As is seen from Fig. \ref{fig:RRR}, the proof of the
Yang-Baxter equation is reduced to sequential transferring
of vertical lines from the left to the right through intersection
points of the other lines with the use of
the star-triangle relations (\ref{st1a}) and (\ref{st1b})
at each step.
Let us note that the both sides of the
star-triangle relations (\ref{st1a}) and
(\ref{st1b}) represent the kernels of
the difference operators explicitly written
in Appendix B ((\ref{st2a}) and (\ref{st2b})
respectively).

\begin{figure}[t]
   \centering
        \includegraphics[angle=-00,scale=0.50]{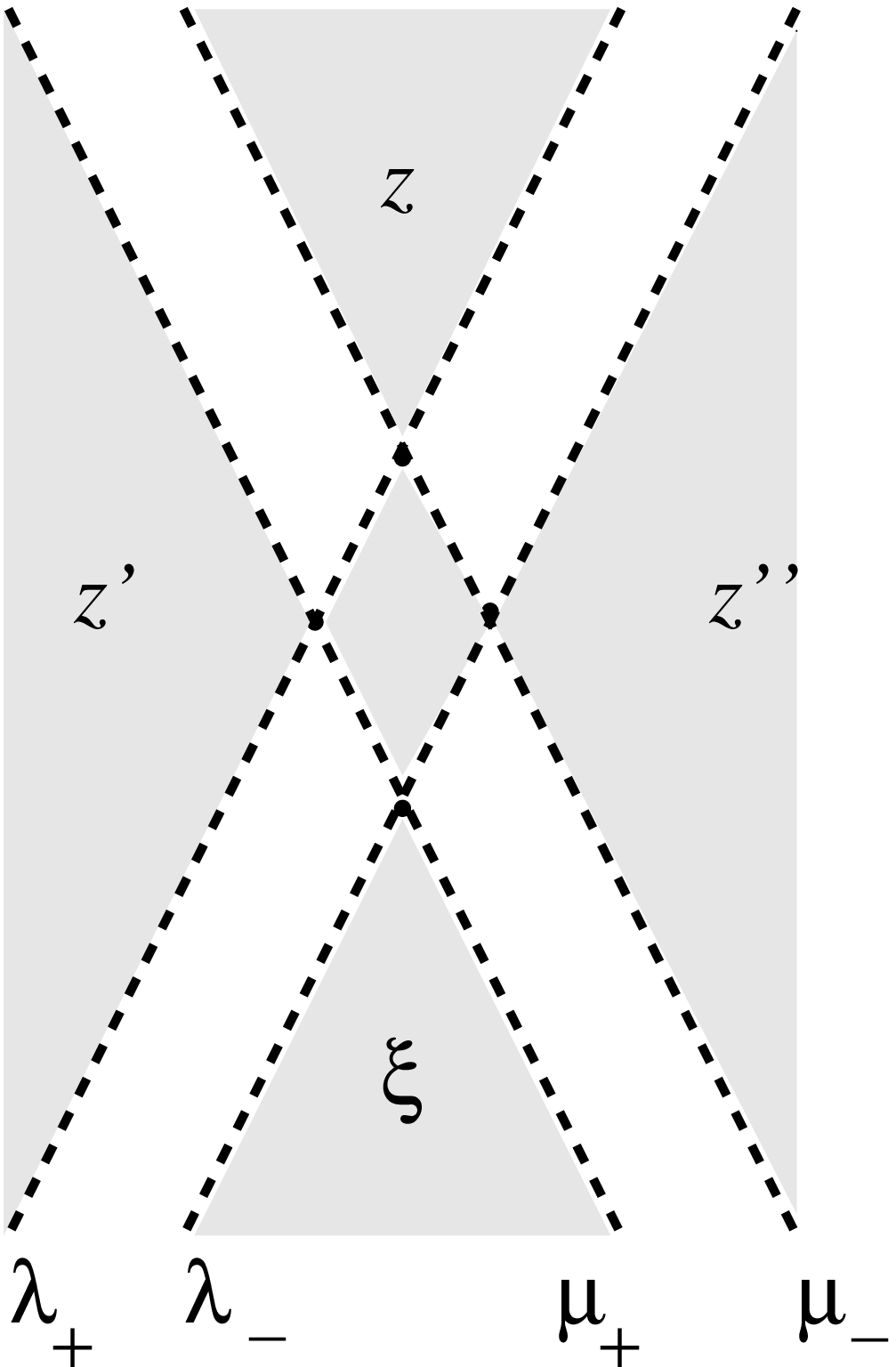}
        \caption{\it The kernel $S^{z}_{\xi}(z',z'')$
        ``dual" to the kernel of the $R$-operator
        (cf. Fig. \ref{fig:R}).}
    \label{fig:S}
\end{figure}

There is an object ``dual" to the
$R$-operator $\check {\sf R}$
in the sense that its kernel is graphically represented by the
same pattern, with shadow parts of the plane being complimentary to those
in Fig. \ref{fig:R}. This duality provides a transformation
which is an infinite-dimensional version of the vertex-face
correspondence.
It sends the $R$-operator to a difference operator
in one variable rather than two. We call it the $S$-operator.
It acts in the variable $z$ and depends on $z'$ and $z''$
as parameters.
Its kernel, $S^{z}_{\xi}(z',z''| \lambda_{+}, \lambda_{-};
\mu _{+}, \mu_{-})$, or simply
$S^{z}_{\xi}(z',z'')$ in short, is shown in Fig. \ref{fig:S}.
This kernel is to be regarded
as an $R$-matrix for a face-type model
with complex variables associated to shadow parts of the plane.
It generalizes the fused Boltzmann weights of the SOS-type
8-vertex model \cite{DJMO}.
According to Fig. \ref{fig:S}
it reads
\beq\label{S1}
S^{z}_{\xi}(z',z'')=\int \! d\zeta
W^{z}_{\zeta}(\lambda_{+}-\mu_{-})
W^{z', \zeta}(\lambda_{+}-\mu_{+})
W^{\zeta , z''}(\lambda_{-}-\mu_{-})
W^{\zeta}_{\xi}(\lambda_{-}-\mu_{+}).
\eeq
The convolution is taken with respect to the variable
sitting in the finite parallelogram at the center of
Fig. \ref{fig:S}. Since each of the two $W^{z}_{\zeta}$-vertices
is represented by a half-infinite sum of the type (\ref{intw8}),
the whole expression (\ref{S1}) is a double sum.
Performing the convolution and re-arranging the double sum, we can
write
$$
S^{z}_{\xi}(z',z'')=c(\lambda_{+}\! -\! \mu_{-})c(\lambda_{-}\! -\! \mu_{+})
\theta_1(2\xi)\sum_{n\geq 0} A_n(z, z', z'')\,
\delta (z\! -\! \xi \! -\! \lambda_{+}\! -\!
\lambda_{-}\! +\! \mu_{+}\! +\! \mu_{-}\! +\! 2n\eta ),
$$
where
$$
\begin{array}{ll}
& A_n(z, z', z'')
\\ &\\
=& \displaystyle{\sum_{k=0}^{n}
\frac{\theta_1(2z\! -\! 2\lambda_{+}\! +\! 2\mu_{-}\! +\! 4k\eta )\,
W^{z', z-\lambda_{+} +\mu_{-}+2k\eta}(\lambda_{+}\! -\! \mu_{+})\,
W^{z-\lambda_{+} +\mu_{-}+2k\eta, z''}
(\lambda_{-}\! -\! \mu_{-})}{W^{z-\lambda_{+} +\mu_{-}+2k\eta, z}
(\lambda_{+}\! -\! \mu_{-}\! +\! \eta )\,
W^{z-\lambda_{+}-\lambda_{-}+\mu_{+}+\mu_{-}+2n\eta ,\,
z-\lambda_{+} +\mu_{-}+2k\eta}
(\lambda_{-}\! -\! \mu_{+}\! +\! \eta )}}.
\end{array}
$$
Using the explicit form of the $W$-functions
it is straightforward to show
that the kernel $S^{z}_{\xi}(z',z'')$ is expressed in terms of the elliptic
hypergeometric series ${}_{10}\omega _{9}$ as follows:
\beq\label{S2}
\begin{array}{lll}
S^{z}_{\xi}(z',z'')&=&
\displaystyle{C \theta_1(2\xi)
e^{2\pi i \xi +\frac{2\pi i}{\eta}((\lambda_{+}-\lambda_{-})z
+(\lambda_{-}-\mu_{+})\xi +(\mu_{+}-\lambda_{+})z')}}
\\ &&\\
&\times &\displaystyle{
\frac{\tilde \theta_1 (z-z' +\mu_{-}-\mu_{+}+\eta )\,
\tilde \theta_1 (z-\xi +\mu_{+}+\mu_{-}-
\lambda_{+}-\lambda_{-})}{\tilde \theta_1
(z\! -\! z' \! +\! \mu_{-}\! +\! \mu_{+}\! -\!
2\lambda_{+} \! +\! \eta )\,
\tilde \theta_1 (z\! -\! \xi \! -\! \mu_{+}\! +\!
\mu_{-} \! -\! \lambda_{+}\! +\! \lambda_{-})}}
\\ &&\\
&\times &\displaystyle{
\frac{\G (2z+2\mu_{-}-2\lambda_{+}+2\eta )}{\G
(2z+2\eta )}
\prod_{j=5}^{10} \frac{\G (2 \alpha_j \eta)}{\G
(2 (\alpha_1 \! -\! \alpha_j \! +\! 1)\eta)}}
\\ &&\\
&\times &\displaystyle{
{}_{10}\omega _{9}(\alpha_1; \alpha_4 , \ldots , \alpha_{10})
\sum_{n\geq 0}\delta (z\! -\! \xi \! -\! \lambda_{+}\! -\!
\lambda_{-}\! +\! \mu_{+}\! +\! \mu_{-}\! +\! 2n\eta )}.
\end{array}
\eeq
Here $C$ is a constant which depends on the spectral
parameters, $\tilde \theta_1(x)\equiv
\theta_1 (x|2\eta )$ and the values of the $\alpha_j$'s are
$$
\alpha_1 = \frac{z+\mu_{-}-\lambda_{+}}{\eta},
\quad \alpha_4 =\frac{\mu_{-}-\lambda_{+}}{\eta},
\quad \alpha_{5,6}=\frac{z\pm \xi +\mu_{+}+\mu_{-}-
\lambda_{+}-\lambda_{-}}{2\eta},
$$
$$
\alpha_{7,8}=\frac{z\pm z' +\mu_{-}-\mu_{+}+\eta}{2\eta},
\quad
\alpha_{9,10}=\frac{z\pm z'' +\lambda_{-}-\lambda_{+}+\eta}{2\eta}\,.
$$
One can see that the series ${}_{10}\omega _{9}$ is balanced
(the balancing condition (\ref{mh2}) is
satisfied) and terminating ($\alpha_6 =-n$ because
of the $\delta$-function). Equation (\ref{S2}) is a version
of the Frenkel-Turaev result \cite{FrTur} adopted to
continuous values of parameters and obtained
by a different method.
The $S$-operator satisfies a sort of the Yang-Baxter
equation which can be graphically represented like in
Fig. \ref{fig:RRR} with transparent pieces of the plane
being changed to the
shadow ones and vice versa.

Another object closely related to the $R$-operator
is the ``transfer matrix on 1 site"
\beq\label{S33}
{\sf T}(\lambda_{+}, \lambda_{-}|
\mu_{+}, \mu_{-})=\mbox{tr}_{\mu}
\left ( \check {\sf R}(\lambda_{+}, \lambda_{-}|
\mu_{+}, \mu_{-}){\sf P}\right ),
\eeq
where ${\sf P}$ is the permutation operator
of the two quantum spaces and the trace is taken in
the space associated with the spectral parameters
$\mu_{\pm}$. The kernel of this transfer matrix is
\beq\label{S4}
T^{z}_{\xi}(\lambda_{+}, \lambda_{-}|
\mu_{+}, \mu_{-})= \int d\xi R^{\zeta z}_{\xi \zeta}
(\lambda_{+}, \lambda_{-}|
\mu_{+}, \mu_{-}).
\eeq
It is not difficult to see that this kernel
is expressed
through the kernel $S^{z}_{\xi}(z',z'')$ given by
(\ref{S2}) as follows:
\beq\label{S5}
T^{z}_{\xi}(\lambda_{+}, \lambda_{-}|
\mu_{+}, \mu_{-})= S^{z}_{\xi}(\xi ,z)
(\lambda_{-}, \lambda_{+}|
\mu_{+}, \mu_{-}).
\eeq
(Note the exchange
of the spectral parameters $\lambda_{+}
\leftrightarrow \lambda_{-}$ in the right-hand side.)
The easiest way to see this is to draw the
corresponding pictures.

\section{Concluding remarks}

In this paper we have presented a unified approach to
intertwining operators for quantum integrable models
with elliptic $R$-matrix associated with the Sklyanin algebra.
We work in the most general setting of
infinite-dimensional representations (with
a complex spin parameter $\ell$) realized by difference
operators in the space of functions of a complex variable $z$.
The elementary building blocks are so-called intertwining
vectors and $W$-functions which are defined in terms of their
scalar products. These elements have a nice graphic
representation as diagrams in the transparent/shadow plane
which allows one to easily construct more complicated objects
like $L$-operators, their vacuum vectors and different
kinds of $R$-matrices and to prove relations between them.
An important constituent of the construction is the
intertwining operator for representations with spins
$\ell$ and $-\ell -1$. For general values of $\ell$, it is
given by the elliptic hypergeometric series ${}_{4}\omega_{3}$
with operator argument.

In fact the material presented here is only the very beginning
of the theory of integrable ``spin chains" with elliptic
$R$-matrices and infinite-dimensional space of states at
each site. Indeed, our discussion has been focused on
a single $L$ or $R$ operator which is relevant to a spin chain
of just one site. The next step is to construct the
transfer matrix, i.e., to consider a chain
of the $R$-operators and to take trace in the auxiliary space.
We plan to address this problem elsewhere.
It would be also very desirable to find a direct connection
of our approach with elliptic beta integrals
\cite{Spir10,Spir01}. Presumably, the pairing 
(\ref{inf2}) or (\ref{pairing}) should be replaced by a sum
of residues.

Among other things, the results presented in this paper indicate
convincingly that there should exist a meaningful theory of
infinite-dimensional representations of the Sklyanin algebra.
Such a theory is still to be developed
and this paper may provide some background
in reaching this ambitious goal.

At last, one should keep in mind that the Sklyanin algebra
is just a very particular representative of a wide family
of elliptic algebras \cite{OF} and, moreover, integrable 
systems associated to algebras from 
this class can be constructed \cite{OR}.
It would be very interesting to investigate to what extent
the methods developped in the present paper can be extended
to other elliptic algebras and corresponding integrable models. 
Such an extension will probably require a further generalization
of elliptic hypergeometric series.

\section*{Acknowledgments}
The author is grateful to S.\-Der\-ka\-chov for a discussion
of the work \cite{DKK07}.
This work was supported in part by RFBR grant 08-02-00287,
by joint RFBR grants 09-01-92437-CEa,
09-01-93106-CNRS, 10-01-92104-JSPS
and by
Federal Agency for Science and Innovations of Russian Federation
under contract 14.740.11.0081.

\section*{Appendix A}
\def\theequation{A\arabic{equation}}
\setcounter{equation}{0}

\subsubsection*{Theta-functions}

We use the following definition of the
Jacobi $\theta$-functions:
\beq
\begin{array}{l}
\theta _1(z|\tau)=-\displaystyle{\sum _{k\in \z}}
\exp \left (
\pi i \tau (k+\frac{1}{2})^2 +2\pi i
(z+\frac{1}{2})(k+\frac{1}{2})\right ),
\\ \\
\theta _2(z|\tau)=\displaystyle{\sum _{k\in \z}}
\exp \left (
\pi i \tau (k+\frac{1}{2})^2 +2\pi i
z(k+\frac{1}{2})\right ),
\\ \\
\theta _3(z|\tau)=\displaystyle{\sum _{k\in \z}}
\exp \left (
\pi i \tau k^2 +2\pi i
zk \right ),
\\ \\
\theta _4(z|\tau)=\displaystyle{\sum _{k\in \z}}
\exp \left (
\pi i \tau k^2 +2\pi i
(z+\frac{1}{2})k\right ).
\end{array}
\label{theta}
\eeq
They also can be represented as infinite products.
The infinite product representation for the $\theta_1(z|\tau)$
reads:
\beq
\label{infprod}
\theta_1(z|\tau)=i\,\mbox{exp}\, \Bigl (
\frac{i\pi \tau}{4}-i\pi z\Bigr )
\prod_{k=1}^{\infty}
\Bigl ( 1-e^{2\pi i k\tau }\Bigr )
\Bigl ( 1-e^{2\pi i ((k-1)\tau +z)}\Bigr )
\Bigl ( 1-e^{2\pi i (k\tau -z)}\Bigr ).
\eeq
Throughout the paper we write
$\theta _a(x|\tau)=\theta _a(x)$,
$\theta (z|\frac{\tau}{2})=\bar \theta (z)$.
The transformation properties for shifts by the periods are:
\beq
\label{periods}
\theta_a (x\pm 1)=(-1)^{\delta _{a,1}+\delta _{a,2}}
\theta_a (x)\,,
\;\;\;\;\;
\theta_a (x\pm \tau )=(-1)^{\delta _{a,1}+\delta _{a,4}}
e^{-\pi i \tau \mp 2\pi i x}
\theta_a (x)\,.
\eeq
Under the modular transformation $\tau \to -1/\tau$
the $\theta$-functions behave as follows:
\beq
\label{mod}
\begin{array}{l}
\theta_{1}(z|\tau )=i \,\sqrt{i/\tau} \,
e^{-\pi i z^2/\tau }\theta_{1}(z/\tau |-1/\tau )\,,
\\ \\
\theta_{2}(z|\tau )= \sqrt{i/\tau} \,
e^{-\pi i z^2/\tau }\theta_{4}(z/\tau |-1/\tau )\,,
\\ \\
\theta_{3}(z|\tau )= \sqrt{i/\tau}\,
e^{-\pi i z^2/\tau }\theta_{3}(z/\tau |-1/\tau )\,,
\\ \\
\theta_{4}(z|\tau )= \sqrt{i/\tau} \,
e^{-\pi i z^2/\tau }\theta_{2}(z/\tau |-1/\tau )\,.
\end{array}
\eeq

The identities often used in the computations are
\beq
\begin{array}{l}
\bar \theta _4 (x)\bar \theta _3(y)+\bar
\theta _4 (y)\bar \theta _3(x)=
2\theta _4 (x+y)\theta_4 (x-y),
\\ \\
\bar \theta _4 (x)\bar \theta _3(y)-
\bar \theta _4 (y)\bar \theta _3(x)=
2\theta _1 (x+y)\theta_1 (x-y),
\\ \\
\bar \theta _3 (x)\bar \theta _3(y)+
\bar \theta _4 (y)\bar \theta _4(x)=
2\theta _3 (x+y)\theta_3 (x-y),
\\ \\
\bar \theta _3 (x)\bar \theta _3(y)-
\bar \theta _4 (y)\bar \theta _4(x)=
2\theta _2 (x+y)\theta_2 (x-y),
\end{array}
\label{theta34}
\eeq
\beq\label{Fay}
\begin{array}{ll}
& \theta_1(z-a-d)\theta_1(z-b-c)
\theta_1(a-d)\theta_1(c-b)
\\ &\\
+&\theta_1(z-b-d))\theta_1(z-a-c)\theta_1(b-d)
\theta_1(a-c)
\\ &\\
=&\theta_1(z-c-d)\theta_1(z-a-b)\theta_1(a-b)
\theta_1(c-d).
\end{array}
\eeq

By $\Theta_n$ we denote the space of $\theta$-functions
of order $n$, i.e., entire functions
$F(x)$, $x\in \CC$, such that
\beq
F(x+1)=F(x)\,,
\;\;\;\;\;\;
F(x+\tau)=(-1)^n e^{-\pi i n\tau -2\pi i nx}F(x)\,.
\label{8}
\eeq
It is easy to see that $\mbox{dim} \,\Theta_n =n$.
Let $F(x)\in \Theta_n$, then $F(x)$ has a multiplicative
representation of the form
$F(x)=c\prod _{i=1}^{n}\theta_1(x-x_i)$,
$\sum _{i=1}^{n}x_i =0$,
where $c$ is a constant. Imposing, in addition to (\ref{8}),
the condition $F(-x)=F(x)$, we define the space
$\Theta_{n}^{+}\subset \Theta_{n}$ of {\it even}
$\theta$-functions of order $n$, which
plays the important role in representations
of the Sklyanin algebra. If $n$ is an even number,
then $\mbox{dim}\, \Theta_{n}^{+} =\frac{1}{2}n +1$.

\subsubsection*{Elliptic gamma-function}

Here we collect the main formulas on the elliptic
gamma-function \cite{R3,FV3}. We use the (slightly modified)
notation of \cite{FV3}.
The elliptic gamma-function
is defined by the double-infinite product
\beq
\label{gamma}
\Gamma(z|\tau, \tau ')=\prod_{k,k'=0}^{\infty}
\frac{1-e^{2\pi i ((k+1)\tau +(k'+1)\tau ' -z)}}
{1-e^{2\pi i (k\tau +k'\tau ' +z)}}.
\eeq
A sufficient condition for the product to be convergent is
$\mbox{Im}\,\tau >0$, $\mbox{Im}\,\tau' >0$.
We need the following properties of the elliptic gamma-function:
\beq
\label{gamma1}
\Gamma (z+1|\tau , \tau ')=
\Gamma (z|\tau , \tau ')\,,
\eeq
\beq
\label{gamma2}
\Gamma (z+\tau |\tau , \tau ')=
-ie^{-\frac{\pi i \tau'}{6}}\eta_{D}^{-1}(\tau')e^{\pi i z}
\theta_{1}(z|\tau')
\Gamma (z|\tau , \tau ')\,,
\eeq
\beq
\label{gamma3}
\Gamma (z+\tau' |\tau , \tau ')=
-ie^{-\frac{\pi i \tau}{6}}\eta_{D}^{-1}(\tau)e^{\pi i z}
\theta_{1}(z|\tau)
\Gamma (z|\tau , \tau ')\,,
\eeq
where
$$
\eta_{D}(\tau)=e^{\frac{\pi i \tau}{12}}
\prod_{k=1}^{\infty}\Bigl ( 1-e^{2\pi i k\tau}\Bigr )
$$
is the Dedekind function. Another useful property is
\beq
\label{gamma5}
\Gamma (z|\tau , \tau ')
\Gamma (\tau ' -z|\tau , \tau ')=
\frac{ie^{\pi i \tau '/6}\eta_{D}(\tau ')}{e^{\pi i z}
\theta_{1}(z|\tau ')}\,.
\eeq
Note also that $\Gamma (z|\tau , \tau ')
\Gamma (\tau +\tau ' -z|\tau , \tau ')=1$.

Under the modular transformation $\tau \to -1/\tau$ the
elliptic gamma-function behaves as follows \cite{FV3}:
\beq
\label{modg}
\Gamma (z|\tau , \tau ')=e^{i \pi P(z)} \,
\frac{\Gamma (z/\tau \,|-1/\tau , \tau '/\tau )}
{\Gamma ((z-\tau )/\tau '\,|-\tau /\tau ', -1/\tau ')}\,,
\eeq
where
\beq
\label{polP}
\begin{array}{lll}
P(z)&=& \displaystyle{-\frac{1}{3\tau \tau '}\,z^3
+\frac{\tau +\tau ' -1}{2\tau \tau '}\,z^2
-\frac{\tau^2 +\tau '^2 +3\tau \tau ' -3\tau -3\tau ' +1}
{6\tau \tau '}\, z\,-} \\ &&\\
&-&\displaystyle{\frac{(\tau +\tau ' -1)(\tau +\tau ' -\tau \tau ')}
{12 \tau \tau '}}\,.
\end{array}
\eeq

Let us list the most frequently used
formulas for
$\G (z) \equiv \Gamma (z|\tau , 2\eta )$.
Using (\ref{gamma3}) several times, we obtain:
\beq
\label{gam6}
\frac{\G (x+2k\eta )}{\G (x)}=
e^{\pi i \eta k^2}R^{-k}e^{\pi i kx}
\prod_{j=0}^{k-1}\theta_1 (x+2j\eta )\,,
\eeq
\beq
\label{gam7}
\frac{\G (x-2k\eta )}{\G (x)}=
(-1)^k e^{\pi i \eta k^2}R^{k}e^{-\pi i kx}
\prod_{j=0}^{k-1}
\Bigl ( \theta_1 (-x+2\eta +2j\eta )\Bigr )^{-1}\,,
\eeq
where $R=ie^{\pi i (\eta +\tau /6)}\eta_{D}(\tau )$.
In particular, ratios of such functions are expressed
through the elliptic Pochhammer symbols as
\beq\label{gam6a}
\begin{array}{l}
\displaystyle{
\frac{\G (2\alpha \eta + 2 k\eta )}{\G (2\beta \eta + 2 k\eta )}
=e^{2\pi i (\alpha -\beta )k\eta}\,
\frac{\G (2\alpha \eta )}{\G (2\beta \eta )}\,
\frac{[\alpha ]_k}{[\beta ]_k}},
\\ \\
\displaystyle{
\frac{\G (2\alpha \eta - 2 k\eta )}{\G (2\beta \eta - 2 k\eta )}
=e^{-2\pi i (\alpha -\beta )k\eta}\,
\frac{\G (2\alpha \eta )}{\G (2\beta \eta )}\,
\frac{[1-\beta ]_k}{[1-\alpha ]_k}}\,.
\end{array}
\eeq

As is seen from (\ref{gamma}), the function
$\Gamma (z|\tau , 2\eta )$ has zeros at the points
$z= 2(k+1)\eta + (m+1)\tau +n$,
and simple poles at the points
$z=-2k\eta -m\tau +n$,
where $k,m$ run over non-negative integers
and $n$ over all integers.
The residues of the elliptic gamma-function at the poles
at $z=-2k\eta$, $k=0,1,2, \ldots$ are:
\beq
\label{residue}
\mbox{res}\,\Bigl |_{z=-2k\eta}
\G (z)=(-1)^k
e^{\pi i \eta k^2}R^{k}r_0
\prod_{j=1}^{k}
\Bigl ( \theta_1 (2j\eta )\Bigr )^{-1}\,,
\eeq
where
$$
r_0=
\mbox{res}\,\Bigl |_{z=0}
\G (z)=
-\frac{e^{\pi i(\tau +2\eta )/12}}{2\pi i
\eta_D(\tau)\eta_D(2\eta )}\,.
$$

\subsubsection*{Elliptic hypergeometric series}

Here we follow \cite{FrTur}.
We define the elliptic Pochhammer symbol (the
shifted elliptic factorial) by
\beq
\label{Pohg} [x]_k \equiv [x] [x+1] \ldots [x+k-1]\,,
\eeq
where $[x]=\theta_1(2x\eta )$ (cf. (\ref{binom})).
By definition, the elliptic hypergeometric series is
\beq
\label{mh1}
{}_{r+1}\omega_{r}(\alpha_1 ; \alpha_4 , \alpha_5 , \ldots ,
\alpha_{r+1};z|2\eta , \tau )=
\! \sum_{k=0}^{\infty}z^k
\frac{[\alpha_1 +2k][\alpha_1 ]_k}{[\alpha_1][k]!}
\prod_{m=1}^{r-2}
\frac{[\alpha_{m+3}]_k}{[\alpha_1 \!-\!\alpha_{m+3}\!+\!1]_k}\,.
\eeq
This is an elliptic analog of the very-well-poised basic
hypergeometric series \cite{GR}.  The series is said to be
{\it balanced} if $z=1$ and
\beq
\label{mh2}
r-5+(r-3)\alpha_1 =2\sum_{m=1}^{r-2}\alpha_{m+3}\,.
\eeq
For a series $\sum_{k\geq 0}c_k$
of the form (\ref{mh1}), the balancing condition
(\ref{mh2}) means that the ratio
$c_{k+1}/c_k$ of the coefficients
is an elliptic function of $k$.
For balanced series (\ref{mh1}), we drop the
argument $z=1$ and the parameters $\eta , \tau$ writing
it simply as
${}_{r+1}\omega_{r}(\alpha_1 ; \alpha_4 , \ldots ,
\alpha_{r+1})$. For instance,
\beq
\label{mh3}
{}_{8}\omega_{7}(\alpha_1 ; \alpha_4 , \alpha_5 , \alpha_6 ,
\alpha_7, \alpha_{8})=
\! \sum_{k=0}^{\infty}
\frac{[\alpha_1 +2k][\alpha_1 ]_k}{[\alpha_1][k]!}
\prod_{m=1}^{5}
\frac{[\alpha_{m+3}]_k}{[\alpha_1 \!-\!\alpha_{m+3}\!+\!1]_k}\,.
\eeq

The series is called
{\it terminating} if at least one of the parameters
$\alpha_4 , \ldots , \alpha_{r+1}$ is equal to a negative
integer number. In this case the sum is
finite and there is no problem of convergence.
If, say $\alpha_{r+1}=-n$, then the series terminates at $k=n$.
The terminating balanced
series were shown \cite{FrTur} to possess nice
modular properties.
That is why they were called {\it modular hypergeometric series}.

The modular hypergeometric series obey a number of
impressive identities. One of them is the elliptic analog
of the Jackson summation formula:
\beq\label{Jackson}
\!\! \!\!{}_{8}\omega_{7}(\alpha _1; \alpha_4, \ldots ,
\alpha_7, -n)=\! \frac{[\alpha_1 \! +\! 1]_n
[\alpha_1 \! -\! \alpha_4 \! -\! \alpha_5 \! +\! 1]_n
[\alpha_1 \! -\! \alpha_4 \! -\! \alpha_6 \! +\! 1]_n
[\alpha_1 \! -\! \alpha_5 \!
-\! \alpha_6 \! +\! 1]_n}{[\alpha_1 \! -\! \alpha_4 \! +\! 1]_n
[\alpha_1 \! -\! \alpha_5 \! +\! 1]_n
[\alpha_1 \! -\! \alpha_6 \! +\! 1]_n
[\alpha_1 \! -\! \alpha_4 \! -\! \alpha_5
\! -\! \alpha_6 \! +\! 1]_n}
\eeq
which is valid provided that the balancing condition
$2\alpha_1 +1 = \alpha_4 + \alpha_5 + \alpha_6 + \alpha_7 -n$ is satisfied
(the Frenkel-Turaev summation formula \cite{FrTur}).

A remark on the notation is in order. In the modern notation
\cite{Spir}, what we call ${}_{r+1}\omega _{r}(\alpha_1 ;
\alpha_4 , \ldots , \alpha_{r+1}|\eta , \tau )$
(following \cite{FrTur}), would be
${}_{r+3}V_{r+1}(a_1 ;
a_6, \ldots , a_{r+3}|q^2, p)$ with
$q=e^{2\pi i\eta}$, $p=e^{2\pi i\tau}$,
$a_j =e^{4\pi i\eta\alpha_{j-2}}$. In particular,
our ${}_{4}\omega _{3}$ would be ${}_{6}V_{5}$.
We understand that the modern notation is better justified
by the meaning of the elliptic
very-well-poisedness condition than the old one
and is really convenient in many cases.
However, we decided to use the old Frenkel-Turaev
notation for the reason that the additive parameters
$\alpha_j$ are more convenient for us than
their exponentiated counterparts.
We think that it is simpler than
to introduce a version of
${}_{r+1}V_{r}$ with additive parameters.

\section*{Appendix B}
\def\theequation{B\arabic{equation}}
\setcounter{equation}{0}

In this appendix we give some details of calculations
which involve modular hypergeometric series.

\subsubsection*{The normalization of the $W_{\zeta}^{z}$-kernel}

Let us consider convolution of the kernels
$W_{\zeta}^{z}(\lambda )$ and $W_{z'}^{\zeta}(-\lambda )$
given by equation (\ref{intw8}):
$$
\begin{array}{ll}
&\displaystyle{\int d\zeta W_{\zeta}^{z}(\lambda )
W_{z'}^{\zeta}(-\lambda )}
\\ &\\
=& \displaystyle{\int d\zeta \frac{
c(\lambda )c(-\lambda )\theta_1 (2\zeta )
\theta_1 (2z')}{W^{\zeta ,z}(\lambda +\eta )
W^{z', \zeta}(-\lambda +\eta )}
\sum_{k,k'\geq 0}\delta (z-\zeta -\lambda +2k\eta )
\delta (\zeta -z' + \lambda +2k'\eta )}
\\ &\\
=& \displaystyle{\sum_{k,k'\geq 0}
\frac{c(\lambda )c(-\lambda )\theta_1 (2z-2\lambda +4k\eta )
\theta_1 (2z')}{W^{z-\lambda +2k\eta ,z}(\lambda +\eta )
W^{z', z-\lambda +2k\eta}(-\lambda +\eta )}\,
\delta (z -z' + \lambda +2(k+k')\eta )}
\\ &\\
=& \displaystyle{c(\lambda )c(-\lambda )\sum_{n\geq 0}
\left (\sum_{k=0}^{n} \frac{\theta_1 (2z-2\lambda +4k\eta )
\theta_1 (2z+4n\eta )}{W^{z-\lambda +2k\eta ,z}(\lambda +\eta )
W^{z+2n\eta , z-\lambda +2k\eta}(-\lambda +\eta )}\right )
\delta (z -z' +2n\eta )}\,.
\end{array}
$$
In order to calculate it explicitly, consider the sum
\beq\label{B1}
S_n(z)=\sum_{k=0}^{n} \frac{\theta_1
(2z-2\lambda +4k\eta )}{W^{z-\lambda +2k\eta ,z}(\lambda +\eta )
W^{z+2n\eta , z-\lambda +2k\eta}(\eta -\lambda)},
\eeq
where the $W$-functions are given by (\ref{intw4}):
$$
W^{z-\lambda +2k\eta ,z}(\lambda +\eta )
=e^{-\frac{2\pi i}{\eta}(\lambda +\eta )(z-\lambda +2k\eta )}
\frac{\G (2z+2\eta +2k\eta )
\G (2\eta +2k\eta )}{\G (2z-2\lambda +2k\eta )
\G (-2\lambda +2k\eta )},
$$
$$
W^{z+2n\eta  ,z-\lambda +2k\eta}(\eta -\lambda)
=e^{\frac{2\pi i}{\eta}(\lambda -\eta )(z+2n\eta )}
\frac{\G (2z \! -\! 2\lambda \! +\! 2\eta \! +\! 2n\eta \! +\! 2k\eta )
\G (2\eta +2n\eta -2k\eta )}{\G (2z +2n\eta +2k\eta )
\G (2\lambda +2n\eta -2k\eta )}.
$$
Plugging this into (\ref{B1}) and representing ratios of elliptic
gamma-functions through elliptic Pochhammer symbols
with the help of (\ref{gam6}), (\ref{gam7}), we obtain:
$$
\begin{array}{lll}
S_n (z)&=& \displaystyle{
e^{4\pi i z -\frac{2\pi i}{\eta}\lambda (\lambda +\eta )
-4\pi i (\lambda -\eta )n}}
\\ &&\\
&\times & \displaystyle{\theta_1(2z -2\lambda )\,
\frac{\G (-2\lambda )\G (2z-2\lambda ) \G (2z+2n\eta )
\G (2\lambda +2n\eta )}{\G (2\eta ) \G (2\eta +2n\eta )
\G (2z+2\eta )
\G (2z-2\lambda +2n\eta +2\eta )}}
\\ &&\\
&\times & \displaystyle{
\sum_{k=0}^{n} \frac{[\frac{z-\lambda}{\eta}
+2k][\frac{z-\lambda}{\eta}]_k}{[\frac{z-\lambda}{\eta}][1]_k}\,
\frac{[-\frac{\lambda}{\eta}]_k \,\,
[\frac{z}{\eta}+n]_k [-n]_k}{[\frac{z}{\eta}+1]_k \,\,
[\frac{z-\lambda}{\eta}+n+1]_k [-\frac{\lambda}{\eta}-n+1]_k}}.
\end{array}
$$
The sum in the last line is the terminating balanced elliptic
hypergeometric series
$${}_{8}\omega_{7}\left (\frac{z-\lambda}{\eta};
-\frac{\lambda}{\eta}, \, \frac{z}{\eta}+n,\,
\frac{z-\lambda +\eta}{2\eta},\, \frac{z-\lambda +\eta}{2\eta}, \,
-n\right )$$
which is equal to
$$
\frac{[\frac{z-\lambda}{\eta}+1]_n \,
[1-n]_n \, [\frac{z+\lambda +\eta}{2\eta}]_n \,
[-\frac{z+\lambda + \eta}{2\eta}-n+1]_n}{[\frac{z}{\eta}+1]_n
[-\frac{\lambda}{\eta}-n+1]_n
[\frac{z-\lambda +\eta}{2\eta}]_n
[-\frac{z-\lambda +\eta}{2\eta}-n+1]_n}
$$
(see (\ref{Jackson})). Because of the factor
$[1-n]_n$ this is zero unless $n=0$. Therefore,
$S_n(z)=0$ if $n\geq 1$ and
$$
S_0(z)= e^{4\pi iz -\frac{2\pi i}{\eta}\lambda (\lambda +\eta )}
\, \frac{\G (2\lambda )\G (-2\lambda )
\G (2z)\G (2z-2\lambda ) \theta_1 (2z-2\lambda )}{\G ^2(2\eta )
\G (2z + 2\eta )\G (2z-2\lambda +2\eta )}.
$$
We thus have
$$
\int \! d\zeta \, W_{\zeta}^{z}(\lambda )
W_{z'}^{\zeta}(-\lambda )=
c(\lambda )c(-\lambda )\theta_1(2z)S_0(z)\delta (z-z').
$$
Using identities for the elliptic gamma-function the
product $\theta_1(2z)S_0(z)$ can be simplified to
$$
\theta_1(2z)S_0(z)=\rho_{0}^{-1}e^{-2\pi i \lambda ^2/\eta}
\G (2\lambda )\G (-2\lambda ),
$$
where
\beq\label{rho0}
\rho_0 = \frac{\G (2\eta )}{ie^{\frac{\pi i \tau}{6}}\eta_D (\tau )}=
\frac{e^{\frac{\pi i}{12}(2\eta -3\tau )}}{i\eta_D (2\eta )}.
\eeq
So, setting
\beq\label{B3}
c(\lambda )=
\frac{\rho_0 \, e^{\pi i \lambda ^2/\eta}}{\G (-2\lambda )}
\eeq
we obtain the relation (\ref{intw12}):
$\int \! d\zeta \,  W^{z}_{\zeta}(\lambda )
W^{\zeta}_{z'}(-\lambda )=\delta (z-z')$.

\subsubsection*{The star-triangle relations}

Let us verify the
star-triangle relation (\ref{st1a})
\beq\label{st1}
W^{z',z}(\mu -\nu )W^{z', z''}(\lambda -\mu )
W^{z}_{z''}(\lambda -\nu )=
\int \! d\zeta W^{z}_{\zeta}(\lambda -\mu )
W^{z',\zeta}(\lambda -\nu )W^{\zeta}_{z''}(\mu -\nu )
\eeq
(see Fig. \ref{fig:WWW}). We use formulas (\ref{intw4}),
(\ref{intw8}).
The left hand side is
$$
\begin{array}{ll}
&\displaystyle{c(\lambda -\nu )\theta_1 (2z'')
\frac{W^{z',z}(\mu -\nu )
W^{z', z''}(\lambda -\mu )}{W^{z'', z}(\lambda -\nu +\eta )}
\sum_{n\geq 0}\delta (z-z'' -\lambda +\nu +2n\eta )}
\\ &\\
=&\displaystyle{c(\lambda -\nu )\sum_{n\geq 0}
\theta_1(2z'')
\frac{W^{z',z}(\mu -\nu )
W^{z', z-\lambda +\nu +
2n\eta}(\lambda -\mu )}{W^{z-\lambda +\nu +
2n\eta, z}(\lambda -\nu +\eta )}\,
\delta (z-z'' -\lambda +\nu +2n\eta )}
\\ &\\
= &\displaystyle{c(\lambda -\nu )\sum_{n\geq 0}
C_n (z',z)\delta (z-z'' -\lambda +\nu +2n\eta )},
\end{array}
$$
where $c(\lambda )$ is given by (\ref{B3}) and
$$
\begin{array}{lll}
C_n(z',z)&=&\displaystyle{
e^{-\frac{2\pi i}{\eta}\left [
(\lambda -\nu )z' -(\lambda -\nu +\eta )z+
(\lambda -\nu )(\lambda -\nu +\eta )\right ]}\,
\theta_1(2z-2\lambda +2\nu +4n\eta )}
\\ &&\\
&\times &\displaystyle{
\frac{\G (2\nu -2\lambda )\G (2z-2\lambda +2\nu )
\G (z \! +\! z' \! +\! \mu \! -\! \nu \! +\! \eta )
\G (z'\! -\! z \! +\! 2\lambda \! -\! \mu \! -\! \nu
\! +\! \eta )}{\G (2\eta )\,
\G (2z+2\eta )\, \G (z+z' -2\lambda +\mu +\nu +\eta )\,
\G (z'-z -\mu +\nu +\eta )}}
\\ &&\\
&\times &\displaystyle{
\frac{[\frac{z-\lambda +\nu}{\eta}]_n \,
[\frac{\nu -\lambda}{\eta}]_n \,
[\frac{z+z'+\nu -\mu +\eta}{2\eta}]_n \,
[\frac{z-z'+\nu -\mu +\eta}{2\eta}]_n}{[1]_n
\, [\frac{z}{\eta}+1]_n  \,
[\frac{z+z'-2\lambda +\nu +\mu +\eta}{2\eta}]_n \,
[\frac{z-z'-2\lambda +\nu +\mu +\eta}{2\eta}]_n}}.
\end{array}
$$
One can see from this expression that the left hand side of
(\ref{st1}) is the kernel of the difference operator
\beq\label{st2a}
\begin{array}{ll}
&\displaystyle{e^{\frac{2\pi i}{\eta}(\lambda -\nu )(z-z')
-\frac{\pi i}{\eta}(\lambda -\nu )^2 }
\frac{\G (2z\! -\! 2\lambda \! +\!
2\nu \! +\! 2\eta )
\G (z \! +\! z' \! +\! \mu \! -\! \nu \! +\! \eta )
\G (z'\! -\! z \! +\! 2\lambda \! -\! \mu \! -\! \nu
\! +\! \eta )}{\G (2z+2\eta )\,
\G (z+z' -2\lambda +\mu +\nu +\eta )\,
\G (z'-z -\mu +\nu +\eta )}}
\\ &\\
\times &\displaystyle{
\normord {}_{6}\omega_{5}\left (
\frac{z-\lambda +\nu}{\eta}; \, \frac{\nu -\lambda}{\eta}, \,
\frac{z+z'+\nu -\mu +\eta}{2\eta},\,
\frac{z-z'+\nu -\mu +\eta}{2\eta}; \,
e^{2\eta \p_z} \right )\normord e^{(\nu -\lambda )\p_z}}.
\end{array}
\eeq

Let us turn to the right hand side of (\ref{st1}).
It is
$$
\begin{array}{ll}
&\displaystyle{c(\lambda -\mu ) c(\mu -\nu )
\int \! d\zeta \frac{W^{z',\zeta}(\lambda -\nu )
\theta_1 (2\zeta )\theta_1 (2z'')}{W^{\zeta , z}
(\lambda -\mu +\eta )W^{z'',\zeta}(\mu -\nu +\eta )}}
\\&\\
& \displaystyle{\quad \quad \quad \quad \times
\sum_{k,k'\geq 0}
\delta (z\! -\! \zeta \! -\! \lambda \! +\!
\mu \! +\! 2k\eta )\delta (\zeta \! -\! z'' \! -\! \mu \! +\!
\nu \! +\! 2k'\eta )}
\\&\\
=& \displaystyle{c(\lambda -\mu ) c(\mu -\nu )
\sum_{n\geq 0}B_n (z',z)\,
\delta (z\! -\! z'' \! -\! \lambda \! +\!
\nu \! +\! 2n\eta )},
\end{array}
$$
where
$$
B_n(z',z)=\sum_{k=0}^{n}
\frac{\theta_1(2z\! -\! 2\lambda \! +\! 2\mu \! +4k\eta )
\theta_1(2z\! -\! 2\lambda \! +\! 2\nu \! +4n\eta )\,
W^{z', z-\lambda +\mu +
2k\eta}(\lambda \! -\! \nu )}{W^{z-\lambda +\mu +
2k\eta , z}(\lambda -\mu +\eta )
W^{z-\lambda +\nu +2n\eta , z-\lambda +\mu +
2k\eta}(\mu -\nu +\eta)}.
$$
The next step is to identify this sum with the
terminating elliptic hypergeometric series with a
pre-factor. The latter is essentially a product of ratios
of the $\G$-functions. Specifically, we have:
$$
\begin{array}{lll}
B_n(z',z)&=&\displaystyle{
e^{\frac{2\pi i}{\eta}\left [
(\lambda -\nu )(z'-z)+(\lambda -\mu )(\lambda -\mu +\eta )
+(\lambda -\nu )(\mu -\nu +\eta )\right ]
+4\pi i z +4\pi i (\mu -\nu +\eta )n}}
\\ &&\\
&\times & \displaystyle{
\frac{\G (2\mu \! -\! 2\lambda )\G (2z\! -\! 2\lambda \! +\! 2\mu )
\G (z\! +\! z' \! +\! \mu \! -\! \nu \! +\!
\eta )\G (z'\! -\! z \! +\!
2\lambda \! -\! \mu \! -\! \nu
\! +\! \eta )}{\G (2\eta )\, \G (2z +2\eta )
\G (z\! +\! z' \! -\! 2\lambda \! +\! \mu \! +\! \nu
\! +\! \eta )
\G (z' \! -\! z \! -\! \mu \! +\! \nu \! +\! \eta )}}
\\ &&\\
&\times & \displaystyle{
\frac{\G (2z \! -\! 2\lambda \! +\! 2\nu \! +\! 2n\eta )\,
\G (2\nu \! -\! 2\mu \! +\! 2n\eta )}{\G (2\eta \! +\! 2n\eta )
\G (2z \! -\! 2\lambda \! +\! 2\mu \! +\! 2\eta \! +\! 2n\eta )}\,
\theta_1 (2z -2\lambda +2\mu )}
\\ &&\\
&\times & \displaystyle{
{}_{8}\omega_{7} \left (\alpha _1 ; \alpha _4, \ldots ,
\alpha _7, -n\right )},
\end{array}
$$
where the parameters $\alpha_i$ are:
$$
\alpha_1 = \frac{z-\lambda +\mu}{\eta},\quad
\alpha_4 = \frac{\mu-\lambda}{\eta}, \quad
\alpha_5 = \frac{z-\lambda +\nu}{\eta}+n,
$$
$$
\alpha_6 = \frac{z+z' +\mu -\nu +\eta}{2\eta}, \quad
\alpha_7 = \frac{z-z' +\mu -\nu +\eta}{2\eta}, \quad
\alpha_8 = -n.
$$
The series with these parameters is balanced, so one can
apply the Frenkel-Turaev summation formula (\ref{Jackson}).
The result is
$$
\begin{array}{ll}
&{}_{8}\omega_{7} \left (\alpha _1 ; \alpha _4, \ldots ,
\alpha _7, -n\right )
\\ &\\
=& \displaystyle{\frac{[\frac{z-\lambda +\mu}{\eta}+1]_n \,
[\frac{\lambda -\nu}{\eta}+1-n]_n \,
[\frac{z-z'-\mu +\nu +\eta}{2\eta}]_n \,
[-\frac{z+z'-\mu +\nu +\eta}{2\eta}+1-n]_n}{[\frac{z}{\eta}+1]_n \,
[\frac{\mu -\nu}{\eta}+1-n]_n \,
[\frac{z-z'-2\lambda +\mu +\nu +\eta}{2\eta}]_n \,
[-\frac{z+z'-2\lambda +\mu +\nu +\eta}{2\eta}+1-n]_n}}.
\end{array}
$$
Now it is straightforward to calculate
the ratio $C_n(z', z)/B_n(z', z)$.
One can see that all $z,z'$ and $n$ dependent factors cancel
in the ratio and one is left with
$$
\frac{C_n(z', z)}{B_n(z', z)}
=\frac{\G (2\eta )}{ie^{\frac{\pi i \tau}{6}}\eta _D (\tau )}
\,
\, \frac{e^{\frac{2\pi i}{\eta}(\lambda -\mu )(\nu -\mu )}
\G (2\nu -2\lambda )}{\G (2\mu -2\lambda )
\G (2\nu -2\mu )}=
\frac{c(\lambda -\mu )c(\mu -\nu )}{c(\lambda -\nu )},
$$
where $c(\lambda )$ is given by (\ref{B3}).
This means that the left and right hand sides of
(\ref{st1}) are indeed equal to each other.

The other star-triangle relation, (\ref{st1b}),
is proved in a similar way. We note that its both sides
are kernels of the difference operator
\beq\label{st2b}
\begin{array}{ll}
&\displaystyle{e^{\frac{\pi i}{\eta}(\lambda -\nu )
(2\mu -\lambda -\nu )}
\frac{\G (2z\! -\! 2\lambda \! +\!
2\nu \! +\! 2\eta )
\G (z \! +\! z' \! +\! \lambda \! -\! \mu \! +\! \eta )
\G (z\! -\! z' \! +\! \lambda \! -\! \mu
\! +\! \eta )}{\G (2z+2\eta )\,
\G (z\! +\! z' \! +\! 2\nu \! -\!
\lambda \! -\! \mu \! +\! \eta )\,
\G (z\! -\! z' \! +\! 2\nu \! -\!
\lambda \! -\! \mu \! +\! \eta )}}
\\ &\\
\times &\displaystyle{
\normord {}_{6}\omega_{5}\left (
\frac{z-\lambda +\nu}{\eta}; \, \frac{\nu -\lambda}{\eta}, \,
\frac{z+z'-\lambda +\mu +\eta}{2\eta},\,
\frac{z-z'-\lambda +\mu +\eta}{2\eta}; \,
e^{2\eta \p_z} \right )\normord e^{(\nu -\lambda )\p_z}}.
\end{array}
\eeq

\end{document}